\documentclass[a4paper,11pt]{article}

\pdfoutput=1
\usepackage{jcappub}

\usepackage{xcolor} 
\usepackage{hyperref}
\usepackage{listings}
\usepackage{mathtools}
\usepackage{amsmath}
\usepackage{amsfonts}
\usepackage{amssymb}
\usepackage{bbold}
\usepackage{epsfig}
\usepackage{graphicx}
\usepackage{bm}
\usepackage{array}
\usepackage{hyperref}
\usepackage{listings}
\usepackage{color}
\usepackage{float}
\usepackage[normalem]{ulem}
\usepackage{placeins}
\usepackage{bbold}

\newcommand{\exclude}[1]{}

\usepackage{tikz,xcolor,hyperref}

\definecolor{lime}{HTML}{A6CE39}
\DeclareRobustCommand{\orcidicon}{\hspace{-1mm}
 \begin{tikzpicture}
 \draw[lime, fill=lime] (0,0) 
 circle [radius=0.16] 
 node[white] {{\fontfamily{qag}\selectfont \tiny \,ID}};
 \draw[white, fill=white] (-0.0525,0.095) 
 circle [radius=0.007];
 \end{tikzpicture}
 \hspace{-3mm}
}

\foreach \x in {A, ..., Z}{\expandafter\xdef\csname orcid\x\endcsname{\noexpand\href{https://orcid.org/\csname orcidauthor\x\endcsname}
 {\noexpand\orcidicon}}
}

\title{Neutrino quantum kinetics in three flavors}

\author[a]{Shashank Shalgar\orcidA{}}
\author[a]{and Irene Tamborra\orcidB{}}

\affiliation[a]{Niels Bohr International Academy \& DARK, Niels Bohr Institute,\\University of Copenhagen, Blegdamsvej 17, 2100 Copenhagen, Denmark}

\emailAdd{shashank.shalgar@nbi.ku.dk}
\emailAdd{tamborra@nbi.ku.dk}

\abstract{The impact of neutrino flavor conversion on the supernova mechanism is yet to be fully understood. We present multi-energy and multi-angle solutions of the neutrino quantum kinetic equations in three flavors, without employing any attenuation term for the neutrino self-interaction strength and taking into account neutrino advection and non-forward collisions with the background medium. Flavor evolution is explored within a spherically symmetric shell surrounding the region of neutrino decoupling in the interior of a core-collapse supernova,  relying on the output of a spherically symmetric core-collapse supernova model with a progenitor mass of $18.6 M_\odot$. We select two representative post-bounce times:  $t_{\rm pb} = 0.25$~s (no angular crossings are present and flavor conversion is triggered by slow collective effects) and  $t_{\rm pb} = 1$~s (angular crossings trigger fast flavor instabilities). We find that flavor equipartition is achieved for the late post-bounce time ($t_{\rm pb} = 1$~s), where the (anti)neutrino emission properties among different flavors tend to approach each other. In this case,  $\bar\nu_e$ tends to $\bar\nu_x = (\bar\nu_\mu + \bar\nu_\tau)/2$ and a similar trend  holds for neutrinos. However, flavor equipartition does not occur for our early post-bounce time ($t_{\rm pb} = 0.25$~s). 
Accounting for weak-magnetism corrections,  crossings in the $\mu$ and $\tau$ lepton number angular distributions arise; however, such crossings  have a magnitude smaller  than the one occurring in the electron sector and negligibly affect flavor evolution.  Because of flavor conversion, 
the neutrino heating rate increases up to $30\%$ with respect to the case where flavor conversion is neglected. 
}

\begin{document}

\maketitle

\section{Introduction}
\label{sec:intro}

Neutrino flavor evolution in dense astrophysical environments, such as core-collapse supernovae and neutron-star mergers, is driven by the refraction experienced by neutrinos when passing through a medium of electrons as well as other neutrinos~\cite{1978PhRvD..17.2369W, 1985YaFiz..42.1441M, Mikheev:1986if, Pantaleone:1992eq}. However, unlike neutrino refraction off electrons, neutrino-neutrino self-interaction is a non-linear phenomenon that depends on the neutrino flavor evolving dynamically~\cite{Tamborra:2020cul,Richers:2022zug,Volpe:2023met,Tamborra:2024fcd}. 

Favorable conditions for flavor conversion can be present in the neutrino decoupling region of core-collapse supernovae and neutron-star mergers,  with possible implications on the explosion mechanism and nucleosynthesis~\cite{Ehring:2023abs,Ehring:2023lcd,Mori:2025cke,George:2020veu,Wu:2017qpc,Wu:2017drk,Li:2021vqj,Fernandez:2022yyv}. In particular, while decoupling, neutrinos can undergo fast flavor conversion in the limit of vanishing vacuum frequency (or equivalently for large neutrino number density)~\cite{Sawyer:2005jk, Sawyer:2008zs, Sawyer:2015dsa, Chakraborty:2016lct, Izaguirre:2016gsx}. Fast conversion is possible, if a crossing in the lepton number angular distribution of neutrinos exists~\cite{Chakraborty:2016lct, Izaguirre:2016gsx, Morinaga:2021vmc,Fiorillo:2024bzm}.  In the absence of crossings, the time-scale characterizing collective flavor conversion is longer and defined by a combination of the vacuum oscillation frequency and the neutrino self-interaction strength (slow collective flavor conversion)~\cite{Duan:2005cp, Duan:2006an, Duan:2006jv,Fogli:2007bk, Duan:2010bg,Fiorillo:2023mze,Fiorillo:2025ank}.
 Due to the non-trivial angular distributions of neutrinos in the decoupling region, fast flavor conversion can also be seeded by slow collective effects, leading to an effective crossing in the angular distribution of the neutrino lepton number~\cite{DedinNeto:2023ykt, Shalgar:2024gjt}. Although, the time-scale associated with flavor evolution is shorter than the ones characterizing neutrino advection and non-forward collisions with the background medium, the latter do affect flavor conversion~\cite{Padilla-Gay:2020uxa,Shalgar:2020wcx,Nagakura:2022kic,Kato:2023dcw,Nagakura:2023xhc,Xiong:2024tac,Xiong:2024pue,Liu:2024nku,Fiorillo:2025ank}. 

The numerical solution of the neutrino quantum transport is challenged by the vast gradient of scales entering the problem. Therefore, in order to make the problem tractable, several approximations and/or symmetry assumptions have been employed. One of these being the investigation of neutrino conversion physics in the two-flavor approximation. This simplification was justified because 1.~the atmospheric mass difference, $\Delta m^2$, is larger than the solar one, $\delta m^2$, by a factor of $\mathcal{O}(30)$--and both are much smaller of the neutrino self-interaction strength $\mu \sim \sqrt{2} G_F n_\nu$, with $G_F$ being the Fermi constant and $n_\nu$ the neutrino number density; 2.~the temperature in the core of supernovae is such that the emission properties of $\nu_\mu$ and $\nu_\tau$ can be approximately assumed to be indistinguishable~\cite{Duan:2005cp,Duan:2010bg}. As for slow collective flavor evolution,  the two-flavor approximation was deemed to be a good approximation of the final flavor outcome~\cite{Fogli:2008fj,Dasgupta:2010cd,Dasgupta:2010ae,Dasgupta:2007ws,Dasgupta:2008cd,Friedland:2010sc}. However, in the context of fast flavor conversion,   the off-diagonal component of the density matrix $\rho_{\mu\tau}$ grows faster than $\rho_{e\tau}$ and $\rho_{e\mu}$ in homogenous systems, leading to a different final flavor configuration with respect to the one obtained in the two-flavor solution~\cite{Chakraborty:2019wxe, Shalgar:2021wlj}.

Recent core-collapse supernova and neutron-star-merger simulations show that muon production could take place in the source core, leading to different distributions between the $\mu$ and $\tau$ neutrino flavors~\cite{Bollig:2017lki,Loffredo:2022prq,Ng:2024zve,Pajkos:2024iml}. This finding, together with the fact that the angular distributions of the non-electron flavors may also develop crossings~\cite{Capozzi:2020syn} and that flavor instabilities linked to three flavor effects can largely affect the flavor outcome~\cite{Airen:2018nvp,Chakraborty:2019wxe,Capozzi:2020kge,Shalgar:2021wlj,Grohs:2022fyq,Purcell:2024bim,Xiong:2024tac}, calls for an investigation of the neutrino quantum kinetics in three flavors. In this paper, we do so by expanding our multi-angle and multi-energy solution of the neutrino quantum kinetic equations in two flavors~\cite{Shalgar:2024gjt}; however, for the first time, we do not employ any attenuation for the self-interaction term--this is crucial to assess  the impact of the vacuum term on self-interaction. We assume identical $\nu_\mu$ and $\nu_\tau$ distributions in the absence of flavor conversion and solve the equations of motion of neutrinos in a spherically symmetric shell embedding the region of neutrino decoupling, relying on inputs from a one-dimensional core-collapse supernova model with a mass of $18.6\ M_\odot$~\cite{SNarchive}. Our goal is to identify  differences in the $\nu_\mu$--$\nu_\tau$ sector due to flavor conversion.

This paper is organized as follows. In Sec.~\ref{sec:eom}, the neutrino equations of motion in three flavors are introduced. Section~\ref{sec:setup} outlines the method adopted to solve the neutrino equations of motion adopting the thermodynamic and hydrodynamic properties of selected post-bounce times of our benchmark supernova model. We present our findings on flavor evolution  in Sec.~\ref{sec:QSC}. The impact of crossings in the angular distribution of the muon- and tau-neutrino lepton numbers is discussed in Sec.~\ref{sec:mucrossing}, while we quantify the feedback of three-flavor effects on the supernova heating rate in Sec.~\ref{sec:heating}. Finally, we summarize our findings in Sec.~\ref{sec:conclusions}. In addition, we explore whether it is justified to neglect the matter term in lieu of effective mixing angles in Appendix~\ref{appA}. 
Appendix~\ref{rbins} provides additional details concerning the radial resolution of our simulations.

\section{Neutrino equations of motion}
\label{sec:eom}
The flavor states for neutrinos and antineutrinos can be described relying on the density matrices,  $\rho(\cos\theta,E,r, t)$ and $\bar{\rho}(\cos\theta,E,r,t)$, that are functions of the neutrino emission angle ($\cos\theta$), propagation radius ($r$), energy ($E$), and time ($t$), respectively. The diagonal components of the density matrices represent the occupation numbers of the flavor states, while the off-diagonal components take into account the coherence between the flavor states.

The neutrino flavor evolution is described by the following  equations~\cite{Sigl:1992fn}:
\begin{eqnarray}
\label{timedep1}
i \left( \frac{\partial}{\partial t} + \vec{v} \cdot \nabla \right) \rho(\cos\theta,E,r,t) &=& [H(\cos\theta,E,r,t), \rho(\cos\theta,E,r,t)] +  i\mathcal{C}[\rho, \bar{\rho}]\ , 
\\
\label{timedep2}
i \left( \frac{\partial}{\partial t} + \vec{v} \cdot \nabla \right) \bar{\rho}(\cos\theta,E,r,t) &=& [\bar{H}(\cos\theta,E,r,t), \bar{\rho}(\cos\theta,E,r,t)]  
+ i\bar{\mathcal{C}}[\rho, \bar{\rho}]  \ . 
\end{eqnarray}
On the left hand side of Eqs.~\ref{timedep1} and \ref{timedep2},  the  term $\vec{v} \cdot \nabla = \cos\theta {\partial}/{\partial r} + ({\sin^{2}\theta}/{r}) ({\partial}/{\partial \cos\theta})$ takes into account the motion of (anti)neutrinos; note that $r$ and $t$ are not interchangeable in Eqs.~\ref{timedep1} and \ref{timedep2} in the presence of inhomogeneities. 

On the right hand side of Eqs.~\ref{timedep1} and \ref{timedep2}, the commutator encapsulates all the physics linked to flavor conversion. 
The Hamiltonian contains  the vacuum,  matter, and  self-interaction terms:
\begin{eqnarray}
H(\cos\theta,E,r,t) &=& H_{\textrm{vac}}(E) + H_{\textrm{mat}}(r, t) + H_{\nu\nu}(\cos\theta,r, t)\ , \\ 
\bar{H}(\cos\theta,E,r,t) &=& - H_{\textrm{vac}}(E) + H_{\textrm{mat}}(r, t) + H_{\nu\nu}(\cos\theta,r, t) \ .
\end{eqnarray}
Each of these  terms is defined as
\begin{eqnarray}
H_{\textrm{vac}} &=& U_{\textrm{PMNS}} \frac{\textrm{diag}(M^{2})}{2E} U_{\textrm{PMNS}}^{\dagger}\ , \\
H_{\textrm{mat}} &=& \textrm{diag}(\lambda(r), 0, 0)\ ,\\
		H_{\nu\nu} &=& \int_{-1}^{1} d\cos\theta^{\prime} \int_{0}^{\infty} dE [\rho(\cos\theta^{\prime},E,r,t)-\bar{\rho}(\cos\theta^{\prime},E,r, t)] \times (1-\cos\theta \cos\theta^{\prime}) \label{eq:Hnunu}\ .
\end{eqnarray}
In $H_{\textrm{vac}}$, the diagonal matrix $M^{2}$ is a function of the squared neutrino mass differences ($\delta m^2$ and $\Delta m^2$) and  $U_{\textrm{PMNS}}(\vartheta_{\textrm{V}}^{12}, \vartheta_{V}^{13}, \vartheta_{V}^{23}, \delta_{\mathcal{CP}})$ is the $3 \times 3$ Pontecorvo–Maki–Nakagawa–Sakata mixing matrix, which is a function of the three neutrino mixing angles and the $\mathcal{CP}$-violation phase.  In $H_{\textrm{mat}}$, $\lambda(r) = \sqrt{2}G_{\textrm{F}}n_{e}(r)$ is the matter potential with $n_e$ being the number density of electrons.  The neutrino self-interaction Hamiltonian is given by $H_{\nu\nu}$.

The collision terms appearing on the right hand side of Eqs.~\ref{timedep1} and \ref{timedep2} include the emission, absorption, and the direction-changing non-forward scattering terms~\cite{Sigl:1992fn,1990Ap&SS.165...65R}:
\begin{eqnarray}
\mathcal{C} &=& \mathcal{C}^{\textrm{emit}}(E, r) - \mathcal{C}^{\textrm{absorb}}(E, r)\rho(\cos\theta, E, r, t) \nonumber\\
&-& \mathcal{C}^{\textrm{dir-ch}}(E, r) \rho(\cos\theta, E, r, t) + \frac{\mathcal{C}^{\textrm{dir-ch}}}{2}(r, E)\int_{-1}^{1}\rho(\cos\theta^{\prime}, E, r, t)d\cos\theta^{\prime} \nonumber \\ 
&+& \cos\theta \mathcal{C}^{\textrm{ani}}(r, E) \int_{-1}^{1} d\cos\theta^{\prime} \cos\theta^{\prime} \rho(\cos\theta^{\prime}, E, r, t) \ . 
\end{eqnarray}

\section{Problem setup}
\label{sec:setup}
In the following, unless otherwise specified, we neglect the matter term and instead use small mixing angles of $\vartheta_{\textrm{V}}^{12} = \vartheta_{\textrm{V}}^{13} = \vartheta_{\textrm{V}}^{23} = 10^{-3}$; we refer the interested reader to  Appendix~\ref{appA} for a discussion on the  validity of such a choice. Moreover, we assume $\delta_{\mathcal{CP}} = 0$ for the sake of simplicity, $\delta m^2 = 7.5 \times 10^{-5}$~eV$^2$, and $\Delta m^2 = 2.5  \times 10^{-3}$~eV$^2$~\cite{nufit}. 
 We note that, unlike most  papers on the topic, we do not employ  an artificial attenuation factor $\xi$  to reduce the strength of self-interaction in Eqs.~\ref{timedep1} and \ref{timedep2}. The attenuation factor was introduced in Ref.~\cite{Nagakura:2022kic} to make the numerical calculations feasible. Note, however, that we  employ an attenuation factor in Appendixes~\ref{appA} and \ref{rbins}  to optimize the computational costs of our simulations.

The neutrino flavor evolution depends on the thermodynamic and hydrodynamic properties of the source, which enter the collision terms. 
We adopt a static hydrodynamic background and thermodynamical properties for selected post-bounce times of a one-dimensional hydrodynamic core-collapse supernova simulation. Our benchmark supernova model has a progenitor mass of $18.6 M_{\odot}$, with final proto-neutron star baryonic mass of   $1.6 M_{\odot}$,  and SFHo equation of state~\cite{SNarchive}.  This supernova model does not include muons, and  proto-neutron star convection is taken into account through a mixing-length approximation. The energy-dependent collision terms have been implemented as in Ref.~\cite{Shalgar:2024gjt} (see also Appendix A of Ref.~\cite{Shalgar:2023aca}). 
We stress that we assume the distributions of $\nu_\mu$'s and $\nu_{\tau}$'s to be identical in the absence of flavor conversion and determined by pair production and Bremsstrahlung.

In what follows, we consider two representative post-bounce times: $t_{\rm{pb}} = 0.25$~s and $t_{\rm{pb}} = 1$~s. These two times have been chosen since no electron lepton number  (ELN) crossing appears for $t_{\rm{pb}} = 0.25$~s, and therefore flavor conversion is triggered by slow collective effects; for $t_{\rm{pb}} = 1$~s, an ELN crossing is instead present and flavor conversion is driven by the fast flavor instability.  Since the explosion of our supernova model is  triggered artificially at $t_{\rm pb}\sim 0.3$~s,   $t_{\mathrm{pb}} \gtrsim 0.25$~s ($t_{\mathrm{pb}} \gtrsim 1$~s) should be considered as representative of the accretion phase (proto-neutron star cooling phase).
We solved the quantum kinetic equations for $t_{\rm{pb}} = 0.05$, $0.12$, $0.25$, $0.5$, $0.75$, and $1$~s; for all configurations corresponding to $t_{\rm{pb}} \gtrsim 0.5$~s, we find an ELN crossing and the final flavor configuration has a behavior qualitatively similar to the one obtained for $t_{\rm{pb}} = 1$~s (results not shown here).

Equations~\ref{timedep1} and \ref{timedep2} suggest that, in order to investigate neutrino flavor evolution, we need to solve a boundary problem.  To do so, we discretize the density matrices over the energy, the polar angle, and the radial range: we use $25$ energy bins in the range between $0$ and  $50$~MeV, $75$ angle bins, and  $150$ radial bins. The cosine of the polar angle ($\cos\theta$)  ranges between $-1$ and $1$, corresponding to neutrinos traveling in radially backward and forward directions, respectively. Since neutrino decoupling is not instantaneous, but occurs gradually in an extended spatial region~\cite{Janka:2006fh,Tamborra:2017ubu}, we track the neutrino flavor evolution in a radial range such that neutrinos of all flavors are in thermal equilibrium at the smallest radius ($r_{\textrm{min}}$) and all flavors free-stream at the largest radius ($r_{\textrm{max}}$). 
In order to ensure that neutrinos have a Fermi-Dirac distribution at $r_{\textrm{min}}$ and there is negligible flux in the backward direction at $r_{\textrm{max}}$, we solve the quantum kinetic equations in the radial range of $[22,57]$~km for $t_{\textrm{pb}} = 0.25$~s  and we consider $[16,31]$~km for $t_{\textrm{pb}} = 1$~s. 

The temporal evolution of the equations of motion is carried out using an adaptive step size with absolute and relative tolerances of $10^{-6}$. 
The advective term in the equations of motion involves a derivative with respect to the polar angle and  radius;  we perform such derivative by using the central difference method. We refer the reader to Refs.~\cite{Shalgar:2022rjj, Shalgar:2022lvv,Shalgar:2024gjt} for additional details on the setup of the numerical simulations.

Following the approach introduced in Refs.~\cite{Shalgar:2022rjj, Shalgar:2022lvv}, first we assume $H=\bar{H}=0$ and evolve the system until the neutrino flavor configuration reaches a steady state (classical-steady state). Then, taking into account flavor conversion, the system is further evolved until a quasi-steady-state configuration is achieved (i.e., we solve  Eqs.~\ref{timedep1} and \ref{timedep2} until the diagonal components of the density matrices become independent of time, except for some numerical fluctuations; note that the off-diagonal components of the density matrices never reach this state).

\section{Flavor evolution in three flavors}
\label{sec:QSC}
In this section, we characterize the quasi-steady-state flavor configuration obtained as a consequence of flavor conversion. In order to do that, we explore the growth of the off-diagonal elements of the density matrix, as well as the distributions of neutrinos and antineutrinos in angle and energy obtained by taking into account flavor conversion.

\subsection{Quasi-steady-state flavor configuration}
\label{sec:QSS}
\begin{figure}
\includegraphics[width=0.99\textwidth]{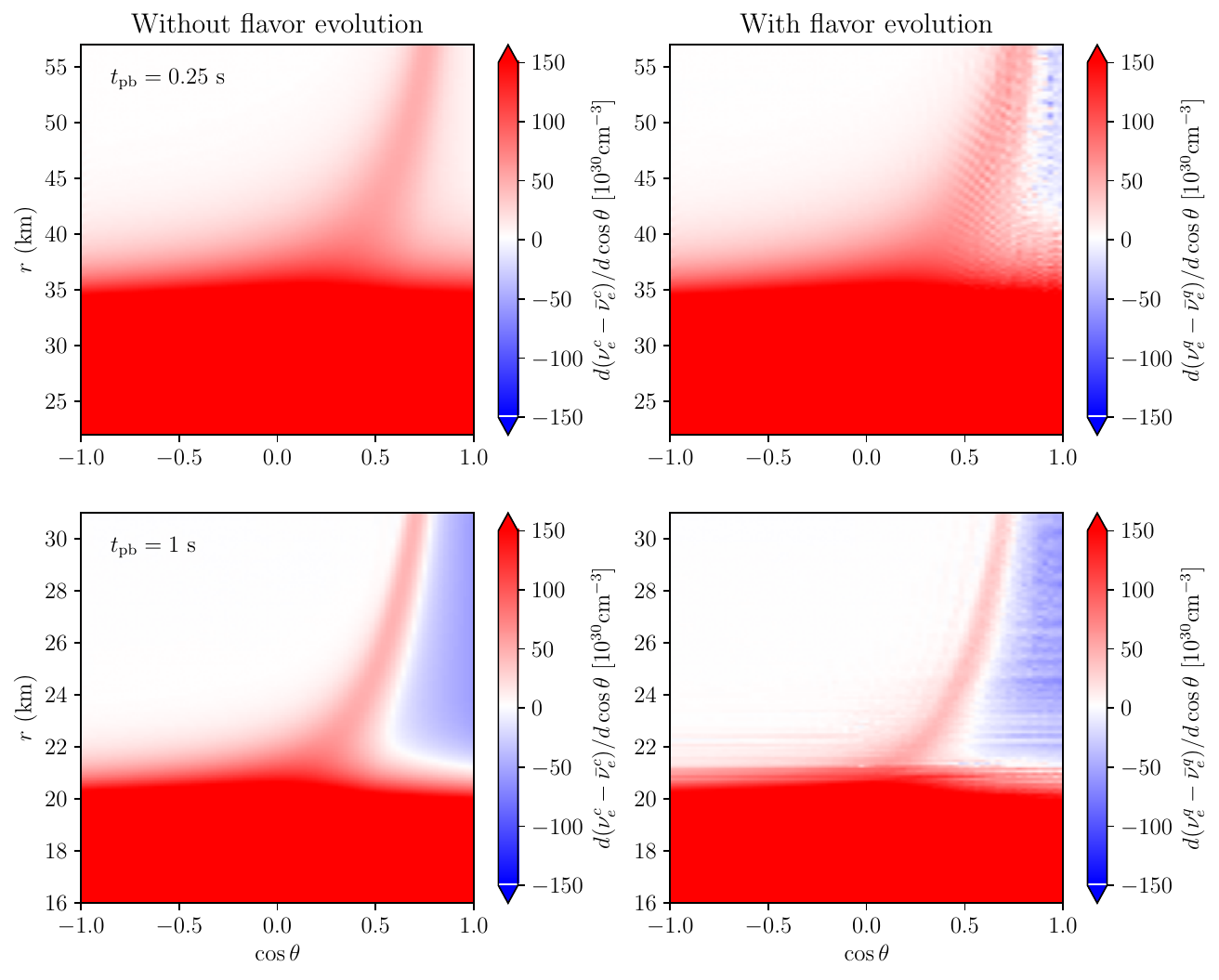}
\caption{Heatmap of  $\rho_{ee} - \bar{\rho}_{ee}$  in the plane spanned by  $\cos\theta$ and $r$ in the absence of flavor evolution (left panels, classical-steady state) and taking into account flavor evolution (right panels, quasi-steady state) for $t_{\textrm{pb}} = 0.25$~s (top panels) and $t_{\textrm{pb}} = 1$~s (bottom panels). For $t_{\textrm{pb}} = 0.25$~s, ELN crossings are not present. Yet, flavor conversion develops triggered by the vacuum terms. For $t_{\textrm{pb}} = 1$~s, ELN crossings develop in the classical-steady-state configuration, with resultant flavor conversion occurring in their proximity.  
}
\label{ELNpanels}
\end{figure}

The left panels of Fig.~\ref{ELNpanels} represent the heatmap of the distribution of $\rho_{ee} - \bar{\rho}_{ee}$ in the plane spanned by  $\cos\theta$ and $r$ for $t_{\mathrm{pb}} = 0.25$~s (top) and $1$~s (bottom) in the absence of neutrino flavor conversion (classical-steady state). As already discussed in Ref.~\cite{Shalgar:2024gjt}, for  $t_{\mathrm{pb}} = 0.25$~s no crossing is present, as signified by the absence of blue colored regions in the heatmap; while  
 the classical-steady-state configuration of $t_{\mathrm{pb}} = 1$~s has crossings in the transition region between the red and blue areas. 

The right panels of  Fig.~\ref{ELNpanels} represent the quasi-steady-state configuration of  $\rho_{ee} - \bar{\rho}_{ee}$ after the three-flavor equations of motion have been evolved for  $t = 2.5 \times 10^{-4}$~s for $t_{\mathrm{pb}} = 0.25$~s and $t = 10^{-4}$~s for $t_{\mathrm{pb}} = 1$~s.
In agreement with the two-flavor analog solution presented in Ref.~\cite{Shalgar:2024gjt} (which however employed an attenuation factor for the self-interaction potential),   significant flavor evolution takes place due to the vacuum term that triggers neutrino self-interactions (slow collective flavor conversion)  for the  $t_{\mathrm{pb}} = 0.25$~s snapshot; note that, although the use of a suppression factor for $H_{\nu\nu}$ may be justified in the context of fast flavor conversion~\cite{Nagakura:2022kic},  it is especially important to avoid  using any attenuation term in the self-interaction Hamiltonian for slow neutrino self-interaction, since the relative strength between $\mu$ and the vacuum terms drives the flavor evolution. On the other hand, for  $t_{\mathrm{pb}} = 1$~s,  flavor conversion develops near the crossing in the $\rho_{ee} - \bar{\rho}_{ee}$ angular distribution (ELN crossing). However, it should be noted that the flavor instability is present only for a small radial region around $20$--$22$~km for $t_{\mathrm{pb}} = 1$~s (cf.~Sec.~\ref{sec:LSA} and Fig.~\ref{timeevol}).
The region beyond $\approx 22$~km displays flavor conversion due to the advection of neutrinos that have undergone flavor evolution at  smaller radii.

\subsection{Growth of the off-diagonal elements of the density matrix}
\label{sec:LSA}
\begin{figure}
\includegraphics[width=0.99\textwidth]{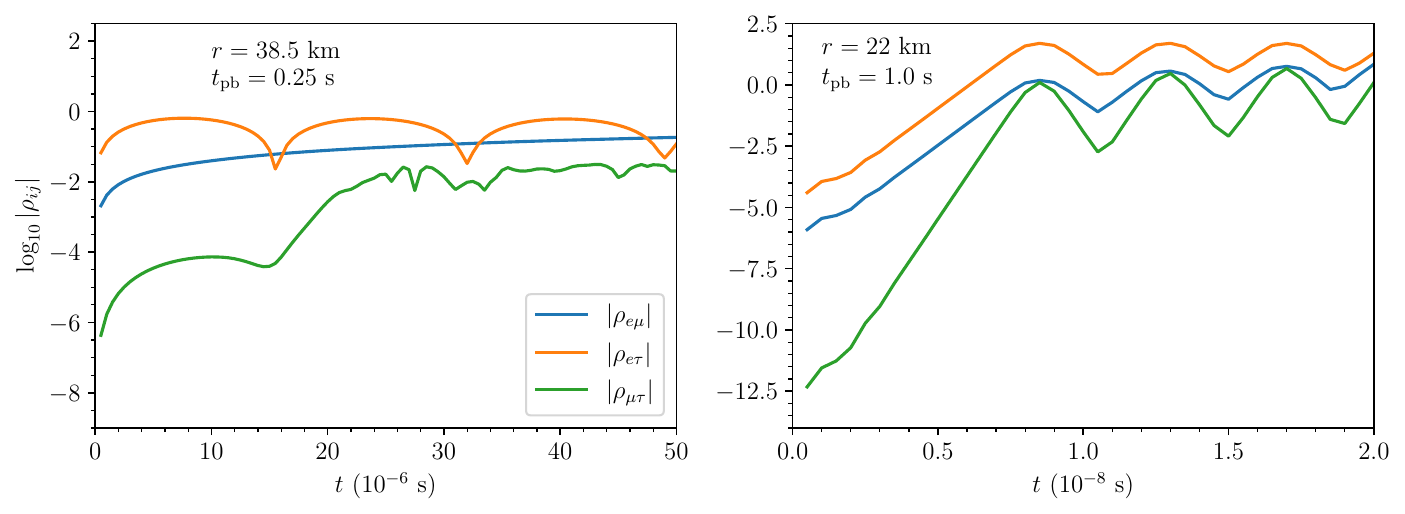}
\caption{Temporal evolution of the off-diagonal components of the density matrix ($\rho_{e\mu}$, $\rho_{e\tau}$, and $\rho_{\mu\tau}$) as functions of time for $t_{\textrm{pb}} = 0.25$~s   ($r= 38.5$~km, left panel) and $t_{\textrm{pb}} = 1$~s  ($r = 22$~km, right panel). The fast growth of $|\rho_{e\mu}-\rho_{e\tau}|$  is responsible for breaking the symmetry between the $\nu_\mu$ and $\nu_\tau$ sectors with fast conversion being triggered by the ELN crossings for $t_{\textrm{pb}} = 1$~s. A slower growth rate is observed for $t_{\textrm{pb}} = 0.25$~s  because the flavor instability is slow and the difference between the $\mu$ and $\tau$ sectors is purely due to the related vacuum terms in the Hamiltonian. Note that the  $x$-scale of the left panel is larger than the one of the right panel.
}
\label{timeevol}
\end{figure}

In order to investigate the role of the $\mu$ and $\tau$ flavors in the flavor evolution, we inspect the temporal evolution of the absolute values of the off-diagonal terms of the density matrices, after energy and angle integration.  Figure~\ref{timeevol} shows the temporal evolution of the off-diagonal terms at selected representative radii falling in the flavor instability region for $t_{\textrm{pb}} = 0.25$~s (left panel) and $t_{\textrm{pb}} = 1$~s (right panel).
For $t_{\textrm{pb}} = 0.25$~s, due to the absence of an ELN crossing, flavor conversion is triggered by the vacuum mixing terms (see also Fig.~6 of Ref.~\cite{Shalgar:2023aca}).  As a consequence, the growth rate of the off-diagonal terms is slower than the one of the fast flavor instability visible in the  right panel of the same figure.
On the other hand, the temporal evolution of the off-diagonal terms of the density matrices for $t_{\textrm{pb}} = 1$~s is driven by the development of the fast instability in the proximity of the ELN crossing. Interestingly, in this case we qualitatively find the same trend observed in the three-flavor studies for homogeneous systems~\cite{Chakraborty:2019wxe, Shalgar:2021wlj}: the off-diagonal component $\rho_{\mu\tau}$ grows faster than $\rho_{e\tau}$ and $\rho_{e\mu}$. The exponential growth of $|\rho_{e\mu}-\rho_{e\tau}|$
is responsible for breaking the symmetry between the $\nu_\mu$ and $\nu_\tau$ sectors~\cite{Shalgar:2021wlj}. Note that, even if  $\rho_{e\mu}$ and $\rho_{e\tau}$ exhibit comparable growth rate, they are seeded by different initial perturbations because of the differences in the related vacuum terms of the Hamiltonian.

\subsection{Three-flavor features of the neutrino distributions in energy and angle}
\begin{figure}
\includegraphics[width=0.99\textwidth]{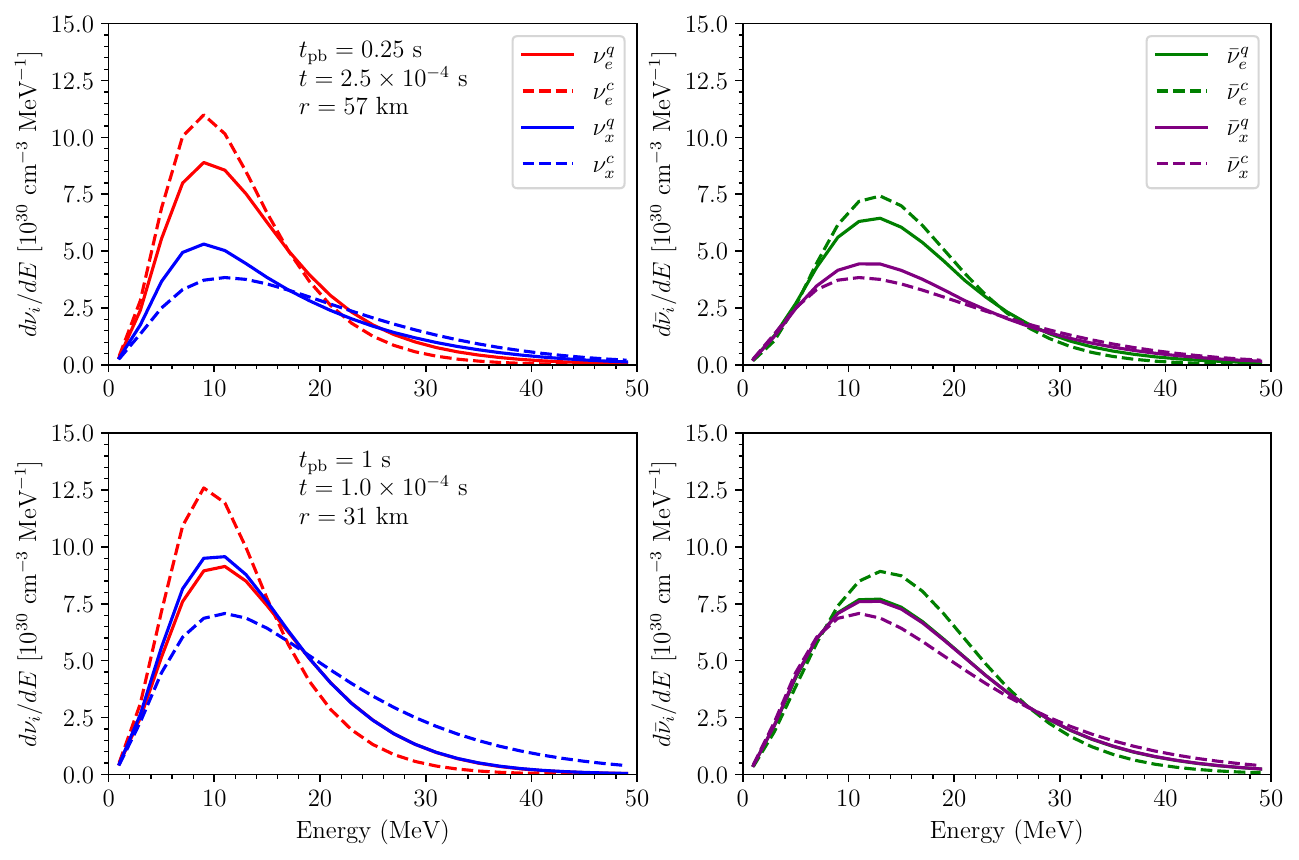}
\caption{Energy spectra for $\nu_{e}$ (red) and $\nu_x$ [blue; with ${\nu}_{x} = ({{\nu}_{\mu} + {\nu}_{\tau}})/{2}$] in the left panels and  $\bar{\nu}_{e}$ (green) and $\bar{\nu}_x$ [purple; with $\bar{\nu}_{x} = ({\bar{\nu}_{\mu} + \bar{\nu}_{\tau}})/{2}$] in the right panels for $t_{\mathrm{pb}} = 0.25$~s on top and $t_{\mathrm{pb}} = 1$~s on the bottom. The energy spectra in the classical-steady-state (quasi-steady-state) configurations are shown as dashed (solid) lines. The quasi-steady-state distributions have been extracted at $57$~km ($31$~km) after  flavor evolution proceeded for $2.5 \times 10^{-4}$~s ($10^{-4}$~s) for $t_{\mathrm{pb}} = 0.25$~s ($t_{\mathrm{pb}} = 1$~s).
Equipartition is   not achieved  $t_{\mathrm{pb}} = 0.25$~s, but it roughly occurs for both  neutrinos and antineutrinos for $t_{\mathrm{pb}} = 1$~s. }
\label{energyspec}
\end{figure}
The differences between the growth rates between the $\nu_\mu$ and $\nu_{\tau}$ sectors propagate into the non-linear phase. Figure~\ref{energyspec} represents the quasi-steady-state energy distributions obtained for  $t_{\mathrm{pb}} = 0.25$~s (top panels) and $1$~s (bottom panels) for neutrinos on the left and antineutrinos on the right.
We find that, for $t_{\mathrm{pb}} = 1$~s, flavor equipartition is achieved for antineutrinos [in the sense that the energy spectrum for $\bar{\nu}_{e}$ becomes comparable with that of $\bar{\nu}_{x} = ({\bar{\nu}_{\mu} + \bar{\nu}_{\tau}})/{2}$] and roughly in the neutrino sector as well. However, this is not the case for  $t_{\mathrm{pb}} = 0.25$~s.
This is due to the fact that,  at late post-bounce times, the neutrino distributions of all flavors tend to approach each other; this, 
in conjunction with the fact that neutrino self-interactions conserve the lepton number, implies that flavor equipartition in the neutrino and antineutrino sectors can be obtained with good approximation. However, this is not possible at earlier post-bounce times, where a larger difference among the neutrino emission properties of all flavors exists.  
It should be noted that, lepton number conservation is not ensured in the numerical solution of the equations of motion because of the collision term. However, we do not find any significant lepton number violation for our benchmark configurations.

It is worth noticing that 
the survival probability in the antineutrino sector is $\sim 1/3$ in the three-flavor scenario when flavor equipartition is achieved (cf.~the bottom panel of Fig.~\ref{energyspec}), while it is expected to be $\sim 1/2$ in the two-flavor case in the case of  equipartition.  However, Fig.~\ref{energyspec} cannot be directly compared with the two-flavor solution presented in Figs.~8 and 9 of Ref.~\cite{Shalgar:2024gjt} since a  suppression factor for the self-interaction strength was adopted in Ref.~\cite{Shalgar:2024gjt}.
 We stress that, although the attenuation factor was used for all post-bounce times in Ref.~\cite{Shalgar:2024gjt} to facilitate the numerical solution of the equations of motions, its use is not appropriate in the absence of ELN crossings (cf. discussion in Sec.~\ref{sec:QSS}).

\begin{figure}
\includegraphics[width=0.99\textwidth]{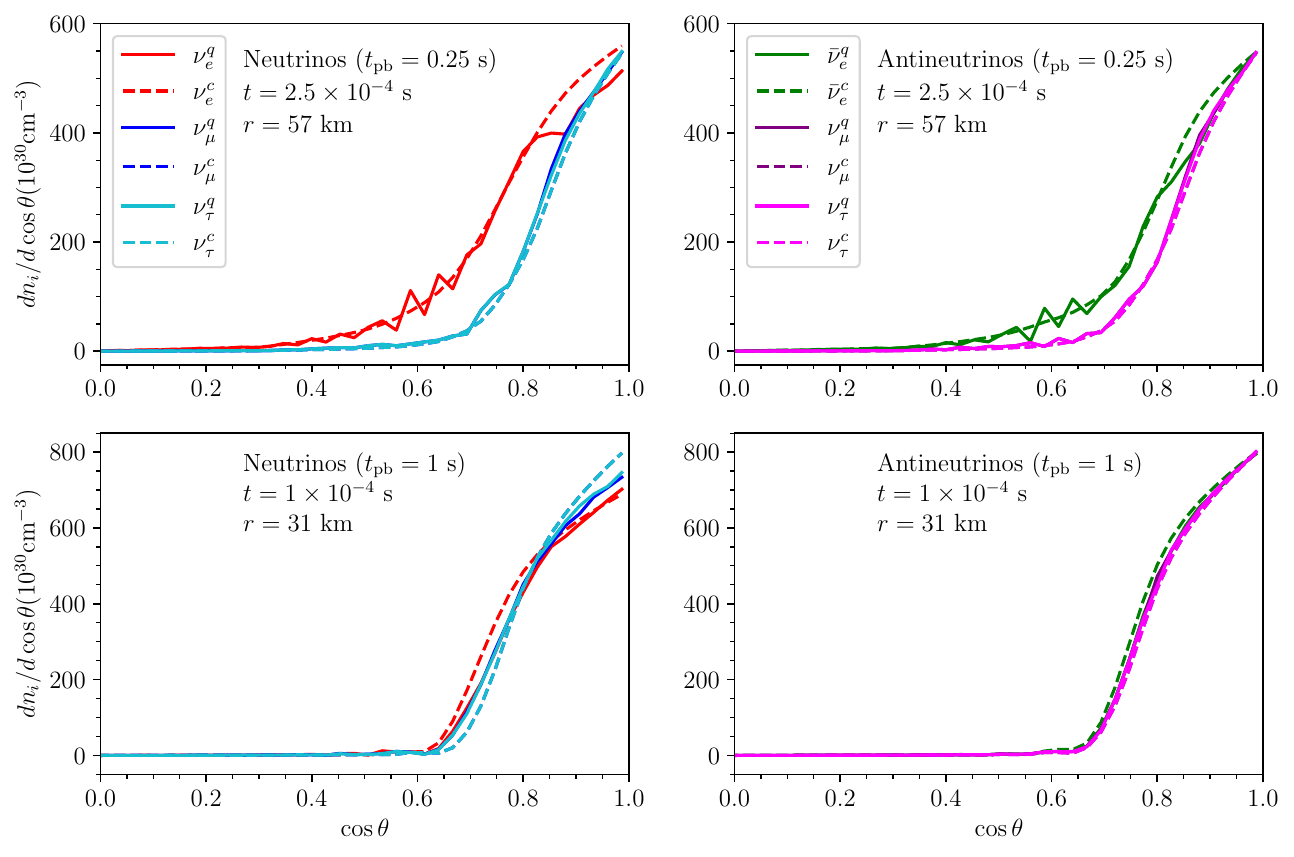}
\caption{Angular distributions for $\nu_{e}$ (red), $\nu_\mu$ (blue), and $\nu_\tau$ (cyan) on the left panels, and for $\bar\nu_{e}$ (green), $\bar\nu_\mu$ (violet), and $\bar\nu_\tau$ (magenta) for $t_{\mathrm{pb}} = 0.25$~s on the top panels and $t_{\mathrm{pb}} = 1$~s   on the bottom. The quasi-steady-state distributions have been extracted at $57$~km ($31$~km) after   flavor evolution occurred for  $2.4 \times 10^{-4}$~s for $t_{\mathrm{pb}} = 0.25$~s ($10^{-4}$~s for $t_{\mathrm{pb}} = 1$~s); cf.~the corresponding energy distributions in Fig.~\ref{energyspec}. 
 The angular distributions in the absence of flavor conversion are shown as dashed lines, whereas the ones obtained including the flavor evolution are shown as solid lines. Note that the $x$-axis ranges between $0$ and $1$ to improve the plot readability, but the angular distributions extend up to $\cos\theta = -1$. The angular distribution of $\nu_{\mu}$ is almost identical to that of $\nu_{\tau}$ as seen from the left panels and the same is true for $\bar{\nu}_{\mu}$ and $\bar{\nu}_{\tau}$ (right panel). 
 }\label{angdist}
\end{figure}

Figure~\ref{angdist} represents the angular distributions associated to the energy spectra in Fig.~\ref{energyspec}. 
For  $t_{\mathrm{pb}} = 0.25$~s, once the quasi-steady-state configuration is achieved, the angular distributions of all  (anti)neutrino flavors remain close to the non-oscillated ones.
On the other hand, except for small differences,  the angular distributions of all flavors of neutrinos and antineutrinos approach each other for $t_{\mathrm{pb}} = 1$~s.  We note that the zig-zag features appearing in the angular distributions of  electron neutrinos and antineutrinos are  plotting artifacts, depending on the  highly oscillatory nature of the solution for $\cos\theta \simeq \mathcal{O}(0.5$--$0.7)$, see Fig.~\ref{ELNpanels}; however, we have tested that the number of particles is conserved in the presence of flavor conversion and we do not find any significant lepton number violation (despite the fact that we take into account collisions). 

\section{Crossings in the muon- and tau-lepton-number distributions}
\label{sec:mucrossing}
Reference~\cite{Bollig:2017lki} pointed out that the large electron chemical potentials and high temperatures characteristic of the proto-neutron star can lead to the creation of muons. The latter could aid neutrino-driven explosions and favor the formation of angular crossings in the muon sector. Inspecting spherically-symmetric core-collapse supernova simulations with and without muons, Ref.~\cite{Capozzi:2020syn} reported the existence of instabilities in the $\mu$--$\tau$ sector due to angular crossings. 

\begin{figure}
\includegraphics[width = 0.99\textwidth]{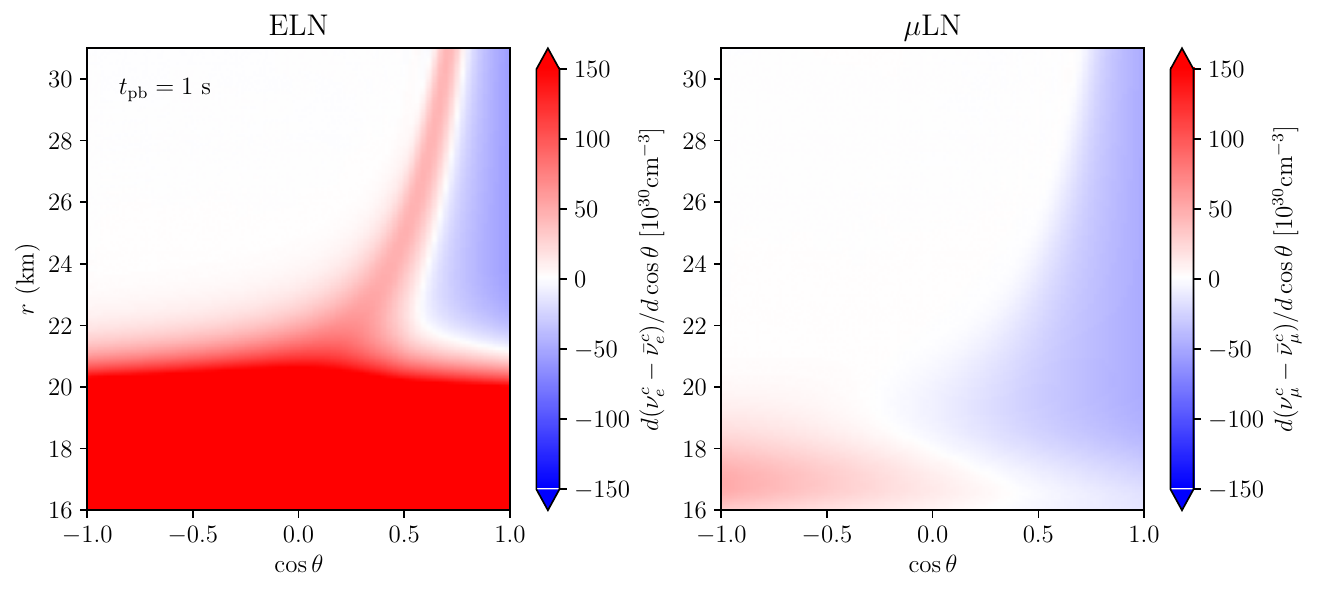}
\caption{{\it Left:} Heatmap of $\rho_{ee}-\bar{\rho}_{ee}$ (ELN distribution) in the plane spanned by $\cos\theta$ and $r$ in the  classical-steady-state configuration for  $t_{\textrm{pb}} = 1$~s (same as the bottom-left panel of Fig.~\ref{ELNpanels}).  {\it Right:} Heatmap of $\rho_{\mu\mu}-\bar{\rho}_{\mu\mu}$  ($\mu$LN distribution) for the same post-bounce time once the weak magnetism corrections are taken into account. The $\tau$LN  distribution is identical to the $\mu$LN one (not shown here). Although  $\mu$LN and $\tau$LN   crossings develop due to the weak magnetism corrections, the non-electron crossings negligibly affect flavor conversion. }
\label{ELNandmuLN}
\end{figure}
Motivated by these findings, we modify our collision term to take into account the weak-magnetism corrections following the recipe presented in Ref.~\cite{thompsonthesis} (note that our supernova model does not include muons, therefore we do not consider muonic  beta reactions).  
For the post-bounce profiles that we have investigated, we find that the collision term is larger for electron flavors due to beta reactions, and therefore the angular distributions are negligibly affected by the weak magnetism corrections; however,  crossings in the muon- and tau-flavor lepton numbers arise due to differences in $\mathcal{C}_{\rm{dir-ch}}$ for $\nu_{\mu/\tau}$ and $\bar\nu_{\mu/\tau}$, as shown in the right panel of Fig.~\ref{ELNandmuLN} (with the angular distribution of $\rho_{\tau\tau}-\bar{\rho}_{\tau\tau}$ being identical to the one of $\rho_{\mu\mu}-\bar{\rho}_{\mu\mu}$). The muon and tau lepton number ($\mu$LN or $\tau$LN) crossings are always much smaller in magnitude than the ELN ones,  as shown in Fig.~\ref{ELNandmuLN}, and they tend to become slightly more prominent at larger radii where the flavor instability due to the ELN has already kicked off. As a consequence, in agreement with the findings of Ref.~\cite{Shalgar:2021wlj} (cf.~their Sec.~III.C),  the growth rate of the flavor instability is driven by the ELN crossing and the $\mu$LN and $\tau$LN  crossings negligibly affect  flavor evolution (results not shown here).

\section{Heating rate}
\label{sec:heating}

In order to quantify the impact of flavor evolution,
we compute the ratio between the neutrino heating rates in the (quasi-)steady-state flavor configuration with and without flavor conversion.
The heating rate,  $\dot{\epsilon} = \dot{\epsilon}_{\nu_{e}} + \dot{\epsilon}_{\bar{\nu}_{e}}$, is defined considering
 \begin{eqnarray}
 \dot{\epsilon}_{\nu_{e}} &=& \sigma_{0} \left(\frac{1+3g_{\textrm{A}}}{4}\right)\int_{0}^{\infty} dE \left(\frac{E+Q}{{m_{e}c^{2}}}\right)^{2} \sqrt{(E+Q)^{2}-m_{e}^{2}}\nonumber\\
 &\times& \left[1-\left(\frac{m_{e}c^{2}}{E+Q}\right)\right]^{\frac{1}{2}}
 \left(1-1.01\frac{E}{m_{n}}\right)\left(1-f_{e^{-}}\right)\frac{d n_{\nu_{e}}}{dE}\ , \\
  \dot{\epsilon}_{\bar{\nu}_{e}}  & = & \sigma_{0} \left(\frac{1+3g_{\textrm{A}}}{4}\right)\int_{m_{e}+Q}^{\infty} dE \left(\frac{E-Q}{{m_{e}c^{2}}}\right)^{2} \sqrt{(E-Q)^{2}-m_{e}^{2}}\nonumber\\
 &\times&
 \left[1-\left(\frac{m_{e}c^{2}}{E-Q}\right)\right]^{\frac{1}{2}}
 \left(1-7.1\frac{E}{m_{p}}\right)\frac{d n_{\bar{\nu}_{e}}}{dE}\ ,
 \end{eqnarray}
where   $Q = 1.2933$~MeV denotes the $Q$-value of the beta reaction, $m_{e}=0.511$~MeV is the mass of the electron, $\sigma_{0}$ is the characteristic neutrino interaction cross section  ($4G_{\textrm{F}}^{2} m_{e}^{2}/\pi \approx 1.7 \times 10^{-44}$~cm$^{2}$)~\cite{thompsonthesis}, $(1-f_{e^{-}})$ is the  Pauli-blocking factor of electrons, and $g_{A}=1.27$ is the axial coupling.

\begin{figure}
\begin{center}
\includegraphics[width=0.75\textwidth]{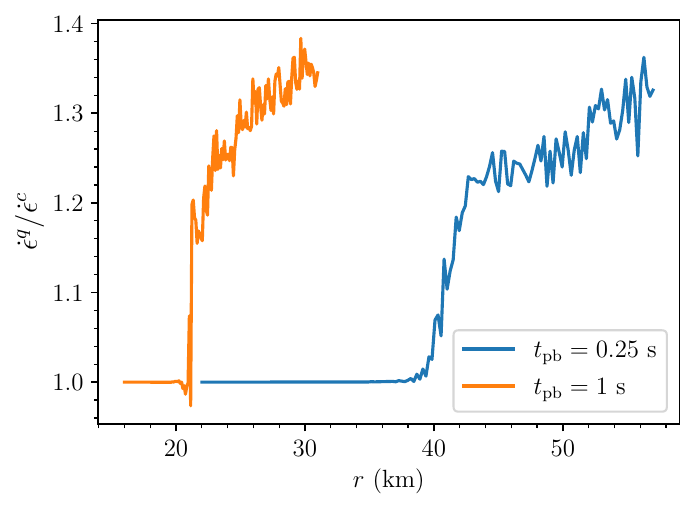}
\caption{Ratio between the neutrino heating rate obtained in the presence of flavor conversion ($\dot{\epsilon}^q$) and without it ($\dot{\epsilon}^c$) when the (quasi-)steady-state flavor configuration is achieved
 for $t_{\textrm{pb}} = 0.25$~s (blue) and $t_{\textrm{pb}} = 1$~s (orange). 
Thanks to neutrino flavor evolution, the electron-type neutrinos acquire a high energy tail due to which the heating rate increases by approximately $30\%$. 
		} 
\label{heating}
\end{center}
\end{figure}

Figure~\ref{heating} shows the radial evolution of the ratio of the heating rates obtained with and without flavor conversion for $t_{\textrm{pb}} = 0.25$~s  and $1$~s. Thanks to flavor evolution, the electron-type neutrinos acquire a high-energy tail; this is a compound effect of the tendency towards flavor equipartition and the fact that the heavy-lepton neutrinos have larger average energies (see also Fig.~\ref{energyspec}).  We find an increase of  $\mathcal{O}(30\%)$ in the heating rate when flavor conversion in three flavors is taken into account. This is due to the fact that neutrino heating is roughly proportional to the third power of neutrino energy.

Previous work, e.g.~Refs.~\cite{Nagakura:2023mhr,Xiong:2024tac}, explored  flavor evolution in a multi-energy and multi-angle framework in three flavors. However,  this work  does not adopt any attenuation term for the self-interaction strength and/or the other terms entering the equations of motion, unlike  assumed in Refs.~\cite{Nagakura:2023mhr,Xiong:2024tac}. A reduction of the heating rate in the gain region was been reported in Refs.~\cite{Nagakura:2023mhr,Xiong:2024tac}. This opposite finding may be related to a larger difference between the  number densities of heavy-lepton and electron flavors than present in our supernova model. As a consequence, a  reduction in number of electron-type neutrinos should be expected throughout the whole energy range (cf.~Fig.~3 of Ref.~\cite{Nagakura:2023mhr}). For our supernova fluid properties, 
the net effect of flavor conversion on the heating rate is dominated by the high-energy tail of the (anti)neutrino spectra.

\section{Conclusions}
\label{sec:conclusions}
Preliminary work in the context of fast flavor conversion~\cite{Chakraborty:2019wxe, Shalgar:2021wlj} concluded that the solution of the quantum kinetic equations in three flavors for a homogeneous system is different from the one expected in two flavors because of the exponential growth of $|\rho_{e\mu}-\rho_{e\tau}|$. 
In this paper, we investigate the flavor solution of a three-flavor neutrino system which is inhomogeneous and non-stationary. In order to do so, we numerically solve the neutrino quantum kinetic equations in a multi-energy and multi-angle framework, relying on static hydrodynamic and thermodynamic inputs from a  $18.6 M_{\odot}$ spherically symmetric core-collapse supernova simulation. Because of the challenges intrinsic to the solution of the quantum kinetic equations, we focus on a spherically symmetric shell in the surroundings of neutrino decoupling. We show our findings for two representative post-bounce times, $t_{\rm pb} = 0.25$ and $1$~s, as examples of cases where  flavor conversion is triggered by slow collective oscillations in the absence of ELN crossings (in the first case) and where flavor conversion is triggered by fast instabilities (for $t_{\rm pb} = 1$~s). First, we neglect flavor conversion and compute the angular distributions of neutrinos and antineutrinos. Then, we adopt such classical-steady-state flavor configuration to solve the quantum kinetic equations in the presence of flavor conversion (without  any attenuation factor for the self-interaction strength), until a quasi-steady-state configuration is reached after   $2.5\times 10^{-4}$~s and $10^{-4}$~s, for $t_{\rm pb} = 0.25$ and $1$~s, respectively.

We find that, for $t_{\mathrm{pb}} = 1$~s, flavor equipartition is achieved in the antineutrino sector, in the sense that the $\bar{\nu}_{e}$ energy spectrum becomes comparable to the one of  $\bar{\nu}_{x} = ({\bar{\nu}_{\mu} + \bar{\nu}_{\tau}})/{2}$; a similar trend is observed for neutrinos. However, flavor equipartition is not found for $t_{\mathrm{pb}} = 0.25$~s.
This difference is due to the fact that, at late post-bounce times, the neutrino emission properties across different flavors tend to approach each other, while a larger difference among the flavor-dependent  emission properties  exists at earlier post-bounce times. 
In agreement with Refs.~\cite{Chakraborty:2019wxe, Shalgar:2021wlj}, an  exponential growth of $|\rho_{e\mu}-\rho_{e\tau}|$ is observed. 
We find that these conclusions  hold for all flavor configurations with an ELN crossing,  corresponding to  post-bounce times $t_{\textrm{pb}} \gtrsim 0.5$~s for our benchmark supernova model. As a consequence of flavor conversion, we observe a  $\mathcal{O}(30\%)$ increase in the heating rate for our  (anti)neutrino configurations.

The weak magnetism corrections to the interaction rates are responsible for the creation of crossings in the angular distributions of the muon and tau neutrino lepton numbers (while they negligibly affect the ELN distributions because beta processes dominate the interaction rate). Such muon/tau crossings have a  magnitude smaller than the ELN ones and give rise to fast instabilities with a smaller growth rate. As a consequence, we find that muon/tau crossings negligibly affect the flavor evolution history.

\acknowledgments
We thank Thomas Janka and Georg Raffelt for helpful discussions. This project has received support from the Villum Foundation (Project No.~13164), the Danmarks Frie Forskningsfond (Project No.~8049-00038B), the European Union (ERC, ANET, Project No.~101087058), and the Deutsche Forschungsgemeinschaft through Sonderforschungbereich SFB 1258 ``Neutrinos and Dark Matter in Astro- and Particle Physics'' (NDM). 
Views and opinions expressed are those of the authors only and do not necessarily reflect those of the European Union or the European Research Council. Neither the European Union nor the granting authority can be held responsible for them. The Tycho supercomputer hosted at the SCIENCE HPC Center at the University of Copenhagen was used for supporting the numerical simulations presented in this work.

\appendix

\section{Matter background and neutrino self-interaction}
\label{appA}

In the neutrino equations of motion,  neutrino refraction on the matter background formally leads to a precession with respect to the interaction direction, similar to the vacuum mixing term. Therefore, within the neutrino-bulb model, refraction of neutrinos off electrons has been neglected, in lieu of a small effective vacuum mixing angle~\cite{Hannestad:2006nj}. Such effective  mixing angle (expected to be smaller as the matter density increases) shifts the onset radius of collective flavor conversion to larger radii~\cite{Hannestad:2006nj,Duan:2007mv,Dasgupta:2010ae,Dasgupta:2011jf}.  
As the matter density increases, one can reach the extreme situation of flavor conversion never being triggered. In fact, early work in this direction concluded that neutrino-neutrino flavor transformation can be suppressed in the core of a core-collapse supernova, during the accretion phase when the electron matter density is significantly larger than the neutrino one~\cite{Esteban-Pretel:2008ovd,Chakraborty:2011gd}.
However, the neutrino angular distributions are not semi-isotropic in the decoupling region, they are forward peaked. 
In addition, within the neutrino-bulb model, the time dependence is ignored in the equations of motion, assuming that the flavor content is in steady state at any time; recent understanding of neutrino flavor conversion as a dynamical phenomenon happening in an inhomogeneous medium highlights the importance of both time and spatial dependence in the solution of the quantum kinetic equations of neutrinos~\cite{Shalgar:2019qwg,Padilla-Gay:2020uxa,Shalgar:2022rjj, Shalgar:2022lvv,Nagakura:2022qko, Nagakura:2022xwe, Xiong:2024tac,Xiong:2024pue,Shalgar:2024gjt,Cornelius:2024zsb}.

In light of these recent developments,  in this appendix, we reexamine the role of dense matter background to understand whether the matter suppression of flavor transformation still holds in a non-stationary neutrino ensemble and for anisotropic neutrino angular distributions. First, we consider the case of the neutrino bulb model and then explore the role of matter in the solution of time- and space-dependent quantum kinetic equations

\subsection{Neutrino bulb model}

In the neutrino-bulb model~\cite{Duan:2006an}, Eqs.~\ref{timedep1} and \ref{timedep2} simplify through the following assumptions. 
\begin{itemize}
\item[-] Neutrino decoupling from matter is modeled assuming that neutrinos are emitted from a sphere of radius $R_{\nu}$ (the neutrinosphere), identical for all flavors  (and not from extended flavor-dependent regions as assumed in this work).  
\item[-]The collision terms in Eqs.~\ref{timedep1} and \ref{timedep2} are neglected ($\mathcal{C}=\bar{\mathcal{C}} = 0$), and neutrinos are emitted semi-isotropically in the outward direction and uniformly from the neutrinosphere.  
\item[-] (Anti)neutrinos are  emitted in pure  flavor eigenstates from the neutrinosphere.
\item[-] The (anti)neutrino ensemble is assumed to be in a steady state (i.e., the time scale of neutrino self-interaction is fast compared to the time scale over which the neutrino density  changes).
\end{itemize}

For a given neutrino trajectory, the angle $\theta$ is a function of $r$ and the emission angle   ($\theta_{0}$) in the bulb model~\cite{Duan:2006an}:
\begin{eqnarray}
\cos\theta = \sqrt{1-\left(\frac{R_{\nu}}{r}\right)^{2}(1-\cos^{2}\theta_{0})}\ .
\end{eqnarray}
Since the (anti)neutrino ensemble is stationary,  time and radius are related for a particular trajectory. At a given location, the time required for  neutrinos to reach a given location depends on the angle relative to the radial direction: 
\begin{eqnarray}
\label{tvsr}
ct &=& r\cos\theta - R_{\nu} \cos\theta_{0}\ , \\
\label{dtvsdr}
c dt &=& dr \cos\theta \ ,
\end{eqnarray}
where we see that the time required to reach a given location is larger for a trajectory with a larger angle $\theta$ with respect to the radial direction. Hence Eqs.~\ref{timedep1} and \ref{timedep2} become dependent on radius (and not time) in the neutrino bulb-model.

The temporal evolution of the density matrix describing the neutrino field along a certain trajectory  is given by
\begin{eqnarray}
\label{rhoevol}
\rho(t) = \exp_{T}\left({-i\int_{0}^{t}}H(E^{\prime},\cos\theta^{\prime},r^{\prime},t^{\prime})dt^{\prime}\right) \rho(0) \exp_{T}\left({i\int_{0}^{t}} H(E^{\prime},\cos\theta^{\prime}, r^{\prime},t^{\prime})dt^{\prime} \right) \ ,
\end{eqnarray}
where the subscript $T$ of the exponential denotes the time ordering. The corresponding radial distance traveled by neutrinos within the time $t$ depends on the angle $\theta$, as illustrated in Eq.~\ref{tvsr}.
If we intend to investigate the  radial evolution of neutrino flavor, it is more convenient to use the following expression:
\begin{eqnarray}
\label{rhoevolr}
\rho(r) = \exp_{T}\left({-i\int_{R_{\nu}}^{r}}H(E^{\prime},\cos\theta^{\prime},r^{\prime})\frac{dr^{\prime}}{c \cos\theta^{\prime}}\right) \rho(R_{\nu}) \exp_{T}\left({i\int_{R_{\nu}}^{r}} H(E^{\prime},\cos\theta^{\prime}, r^{\prime})\frac{dr^{\prime}}{c \cos\theta^{\prime}} \right) \ ,\nonumber \\
\end{eqnarray}
where we use Eq.~\ref{dtvsdr} to account for the longer path length associated with neutrinos with a larger emission angle. It is important to note that the Hamiltonian is rescaled by a factor ${1}/{(c\cos\theta)}$ to consider that neutrinos with a larger emission angle travel a longer distance. 
This led to conclude that matter suppresses flavor conversion in the accretion phase of a core-collapse supernova~\cite{Chakraborty:2011gd}.

\subsection{Matter effects on flavor transformation in the quantum kinetic approach}

We now investigate the role of matter in flavor conversion relaxing the simplifying assumptions characterizing the bulb model and relying on Eqs.~\ref{timedep1} and \ref{timedep2}. This means that we now do not consider a stationary neutrino ensemble. 
In this case, the  motion of neutrinos is taken into account through the additive term (vs.~a multiplicative factor $1/\cos\theta$ appearing in the time-independent formalism). Hence, we should expect a qualitatively different impact of matter on  flavor evolution, since the background matter should suppress flavor instabilities in space, but not in time~\cite{Dasgupta:2016dbv,Abbar:2017pkh}, and a partial cancellation of matter effects occurs for a non-stationary neutrino gas~\cite{Abbar:2015fwa,Dasgupta:2015iia}. 
 
Equations~\ref{timedep1} and \ref{timedep2} include the advective term ($\vec{v} \cdot \nabla$), which is responsible for hindering matter suppression of flavor conversion. 
In fact, if we assume that the advective term suppresses the flavor instability in the presence of a significant matter term, then the term $\vec{v} \cdot \nabla \rho_{ex}$ should be small at the radius where the system transitions from flavor stable to flavor unstable ($r_{\textrm{instab}}$). This implies that the system should behave like a homogeneous gas at $r_{\textrm{instab}}$ and hence be unstable~\cite{Duan:2005cp, Esteban-Pretel:2008ovd}. Matter cannot suppress the neutrino flavor instability due to the advective term. This fact can be demonstrated by performing numerical simulations in the time-dependent formalism with and without the matter term, as shown in the following. 

We focus on the post-bounce time  $t_{\textrm{pb}} = 0.05$~s, which exhibits  flavor instability due to the vacuum term (without ELN crossings). We deliberately choose this configuration because, in the presence of ELN crossings, the growth rate of fast conversion is very large, and matter suppression is not feasible. 
In order to make the problem tractable, we use a suppression factor  $\xi = 10^{-3}$ in Eq.~\ref{eq:Hnunu}; note that this is different than the case considered in the main text (where  no suppression factor is used) in order to minimize the computational time. However, our conclusions should not qualitatively depend on the choice of $\xi$.

We investigate  flavor evolution in three different settings as summarized in Table~\ref{mixparams}.
\begin{itemize}
\item[-] Case A. We assume $\lambda = 0$ in Eqs.~\ref{timedep1} and \ref{timedep2} and small effective mixing angles: $\theta_{12} = \theta_{23} = \theta_{13} = 10^{-3}$~rad. This is the setup adopted in the main text.
\item[-] Case B. We adopt the best-fit values from Ref.~\cite{nufit}  for the mass and mixing parameters, although we assume $\delta_{\rm{CP}}=0$ for the sake of simplicity. We adopt inputs from the supernova simulation to compute the matter potential $\lambda$,  but we suppress it by a constant factor $10^{-2}$ and adopt $\lambda = 10^{3}$~km$^{-1}$ in the innermost radial region, where $\lambda$ would be too large preventing an efficient numerical solution. 
\item[-] Case C. This case is identical to Case B, except for the fact that the matter potential is suppressed by a factor $10^{-3}$ instead of $10^{-2}$.
\end{itemize}
We note that, although the matter Hamiltonian is not as large as it would be in the supernova core, it is still very large compared to the vacuum frequency. Also, unlike the case of the neutrino-bulb model, we have non-trivial angular distributions, which lead to the development of flavor instabilities at a much larger density. Hence, the matter suppression would have been much more pronounced if present.

\begin{table}[]
    \caption{Summary of the three different scenarios adopted to investigate whether matter suppresses flavor conversion in a time-dependent, inhomogeneous neutrino gas. The neutrino mass-mixing parameters are the best-fit values reported in Ref.~\cite{nufit}.}
\label{mixparams}
    \centering
    \begin{tabular}{|l|l|l|l|}
        \hline
        & Case A & Case B & Case C \\
        \hline
        \hline
        $\delta m^{2}$ (eV$^{2})$ & $7.5 \times 10^{-5}$ & $7.5 \times 10^{-5}$ & $7.5 \times 10^{-5}$\\
        \hline
        $\Delta m^{2}$ (eV$^{2})$ & $2.5 \times 10^{-3}$ & $2.5 \times 10^{-3}$ & $2.5 \times 10^{-3}$\\
        \hline
        $\vartheta^{12}_V$ (rad) & $10^{-3}$  & $0.59$  & $0.59$  \\
        \hline
        $\vartheta^{23}_V$ (rad) & $10^{-3}$  & $0.86$  & $0.86$  \\
        \hline
        $\vartheta^{13}_V$ (rad) & $10^{-3}$  & $0.15$  & $0.15$ \\
        \hline
        $\delta_{\mathcal{CP}}$ & 0 & 0 & 0 \\
        \hline
        $\lambda$ (km$^{-1}$) & 0 & min($10^3,10^{-2}\sqrt{2} G_{\textrm{F}} n_{e}$) & min($10^3,10^{-3}\sqrt{2} G_{\textrm{F}} n_{e}$) \\
        \hline
    \end{tabular}
\end{table}

\begin{figure}
\includegraphics[width=0.99\textwidth]{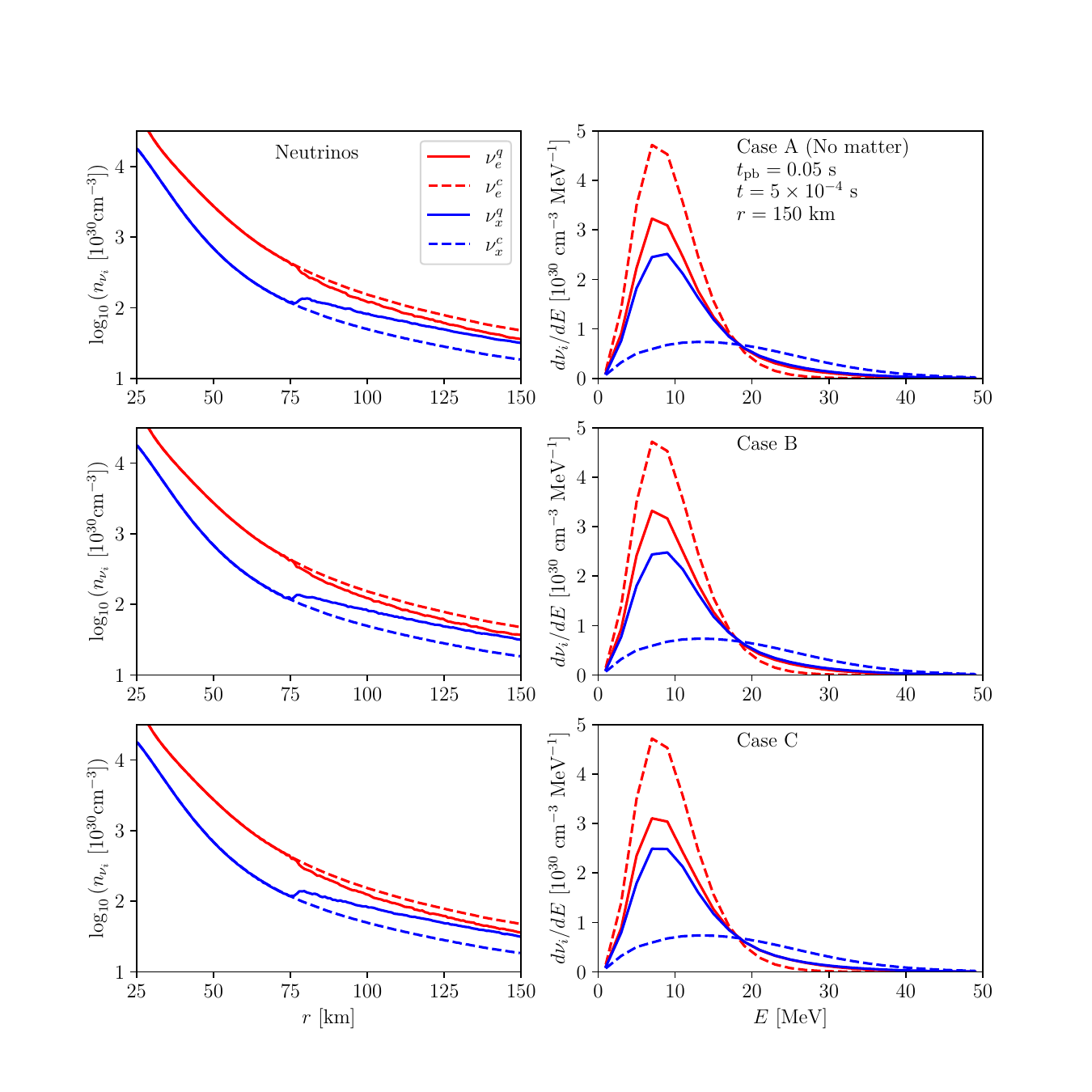}
\caption{{\it Left:} Radial dependence of the neutrino number density for Cases A, B, and C, from top to bottom respectively (cf.~Table~\ref{mixparams}). 
The red solid (dashed) line shows the number density for $\nu_{e}$ with (without) flavor evolution at $t = 5 \times 10^{-4}$~s. The blue solid (dashed) line shows the number density for $\nu_{x}$ with (without) flavor evolution. The number density of $\nu_{x}$ is defined as the average between the  $\nu_{\mu}$ and $\nu_{\tau}$ number densities. {\it Right:} Energy spectra of $\nu_{e}$ and $\nu_{x}$ extracted at $r = 150$~km. The neutrino densities, after flavor conversion, are comparable among Cases A, B, and C.}
\label{fullnu}
\end{figure}
\begin{figure}
\includegraphics[width=0.99\textwidth]{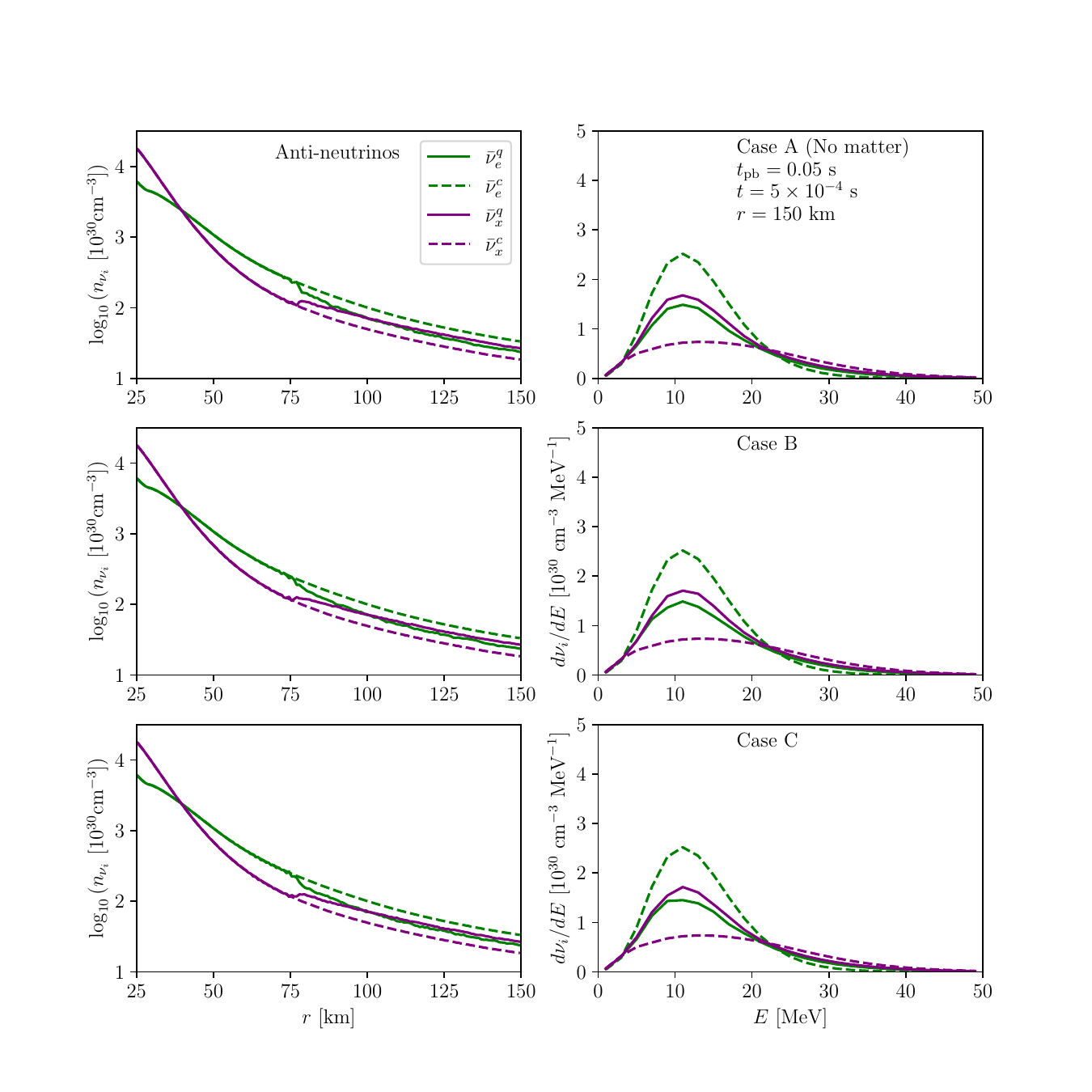}
\caption{Same as Fig.~\ref{fullnu} but for antineutrinos. The $\bar{\nu}_{e}$ [$\bar{\nu}_{x} = (\bar\nu_{\mu}+\bar\nu_{\tau})/2$] number density is plotted in green (purple).}
\label{fullnubar}
\end{figure}

Figures~\ref{fullnu} and \ref{fullnubar} show the results of flavor evolution in the three different Cases at $t=5 \times 10^{-4}$~s. 
 We can see that, irrespective of whether effective mixing angles or $\lambda$ is adopted, the final flavor configuration is almost identical in all three Cases. This highlights that matter suppression does not occur in the presence of large $\lambda$, contrary to what  is expected in the bulb model. Moreover, Fig.~\ref{fullnu} and \ref{fullnubar} also justify the employment of effective mixing angles in numerical simulations of time-dependent, inhomogeneous neutrino gases, instead of the realistic matter potential that would make the numerical solution of the quantum kinetic equations much more expensive.

\section{Dependence of the quasi-steady-state configuration on the number of radial bins}
\label{rbins}
It is important to ensure that the numerical results  are convergent with respect to the number of radial bins. In the main text, the numerical calculations have been performed without any rescaling of the self-interaction strength (cf.~Ref.~\cite{Nagakura:2022kic} for details). However, this choice limits the number of radial bins that can be used in the simulations. The simulations without  the attenuation term are extremely  expensive computationally, they cannot be repeated for a variable number of radial bins. Hence, in this appendix, we show that our results do not depend significantly on the number of radial bins by performing  numerical simulations with a self-interaction strength that is rescaled by a factor of $10^{-2}$.  
\begin{figure}
		\includegraphics[width=0.99\textwidth]{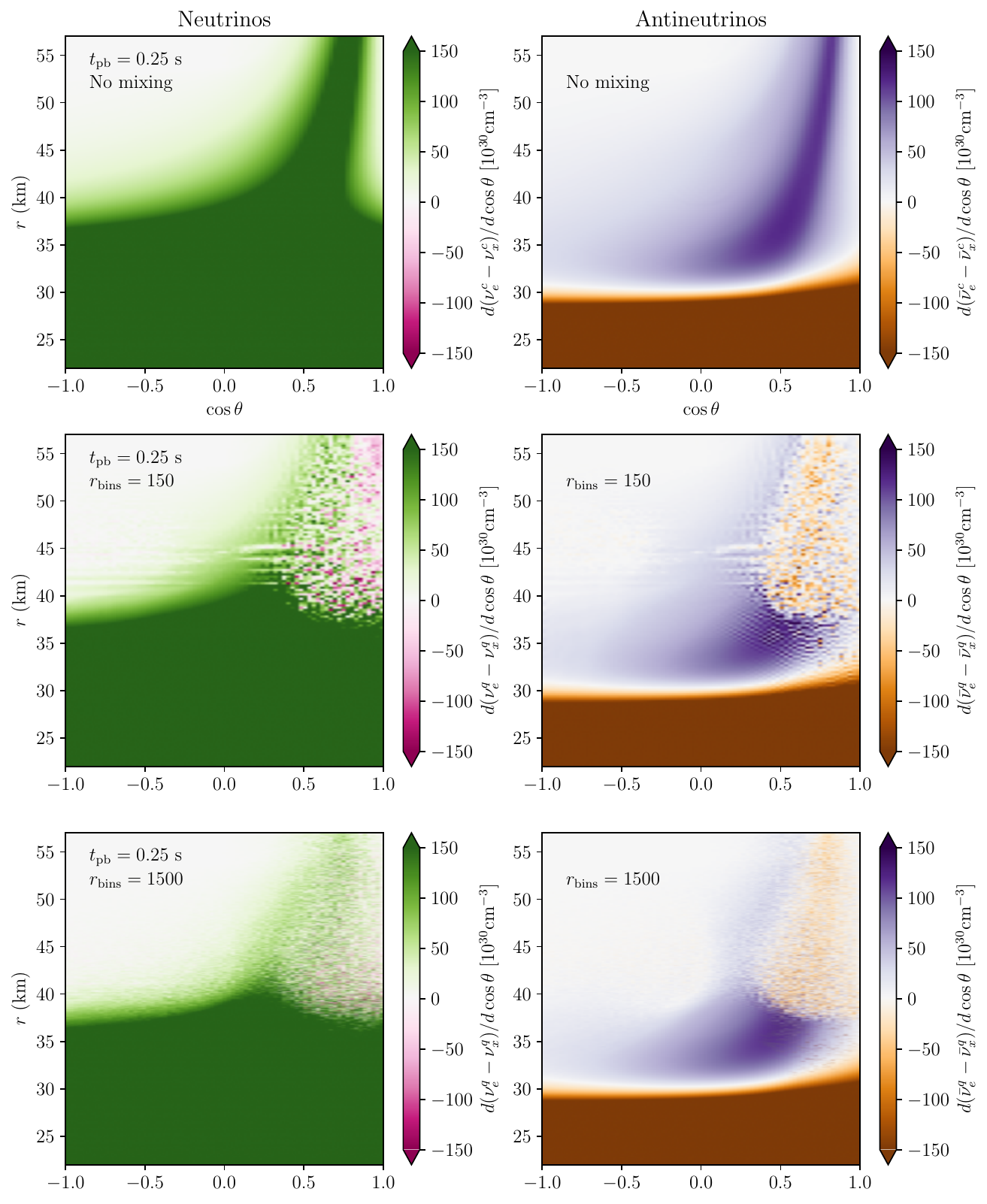}
		\caption{Heatmaps of $\nu_{e}$--$\nu_{x}$ [left panels, with $\nu_{x} = (\nu_{\mu} + \nu_{\tau})/2$] and $\bar{\nu}_{e} - \bar{\nu}_{x}$ (right panels) for $t_{\textrm{pb}} = 0.25$~s, and with the self-interaction term  rescaled by a factor of $10^{-2}$. The top panels show the classical steady-state configuration in the absence of flavor conversion.  The middle (bottom) panels display the quasi-steady-state configurations after $t=10^{-4}$~s, obtained using $150$ ($1500$) radial bins. The quasi-steady-state configurations obtained using $150$ radial bins are nearly identical to the ones obtained using $1500$  bins, implying convergence.}
		\label{tp25heat}
\end{figure}
\begin{figure}
        \includegraphics[width=0.99\textwidth]{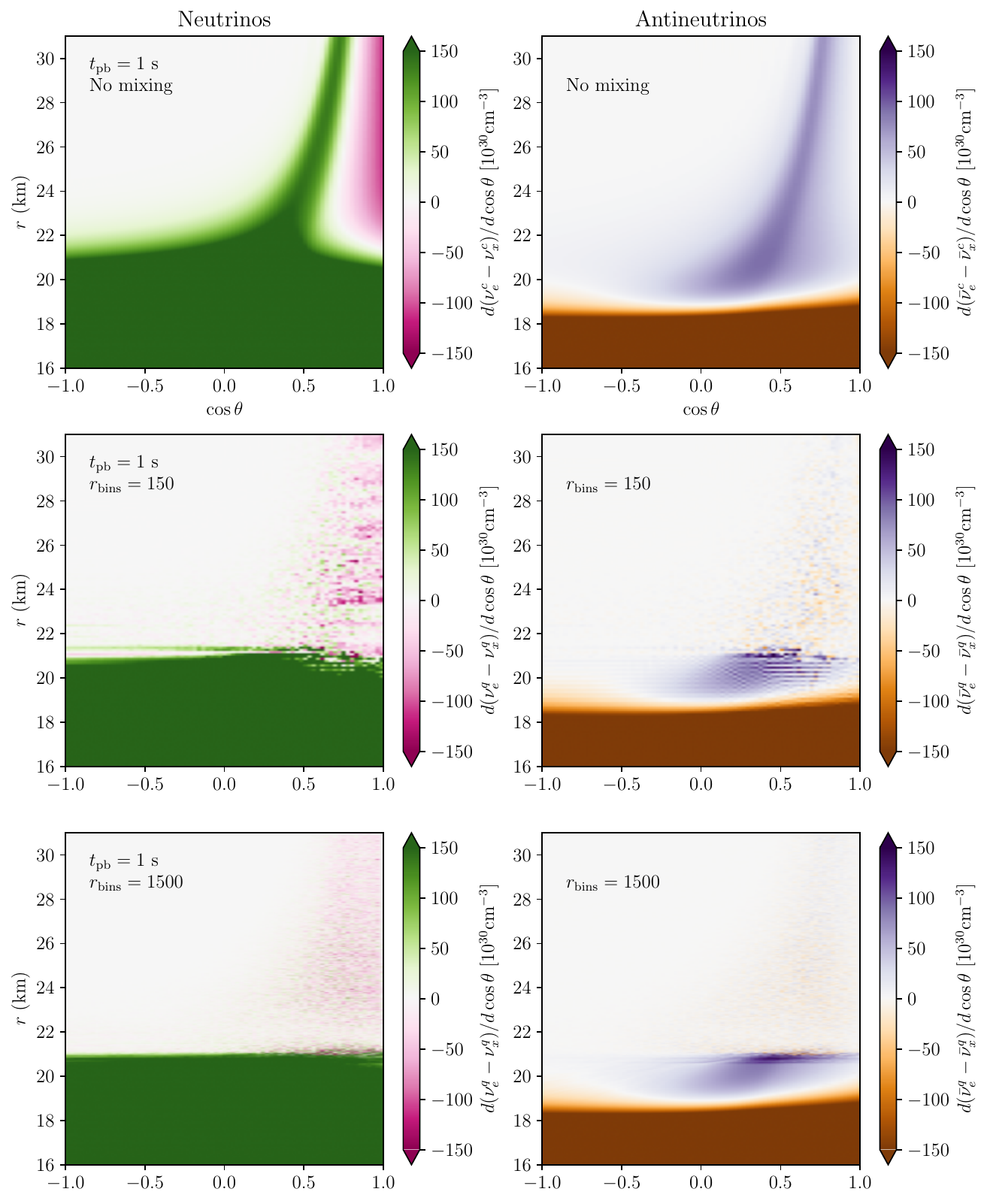}
		\caption{Same as Fig.~\ref{tp25heat}, but for $t_{\textrm{pb}} = 1$~s. Also for this post-bounce time, the quasi-steady-state configurations obtained using $150$ radial bins are nearly identical to the ones obtained using $1500$  bins.}
        \label{t1heat}
\end{figure}

We carry out simulations with  $150$ and $1500$ radial bins for the steady-state flavor configurations at $0.25$~s and $1$~s (except for the attenuation factor for $\mu$, all other simulation inputs remain unchanged).   It should be noted that, since we use an adaptive step size for the temporal steps, the time steps used in this appendix are naturally larger than the ones used in the main text on account of the smaller self-interaction strength.

\begin{figure}[h!]
		\includegraphics[width=0.90\textwidth]{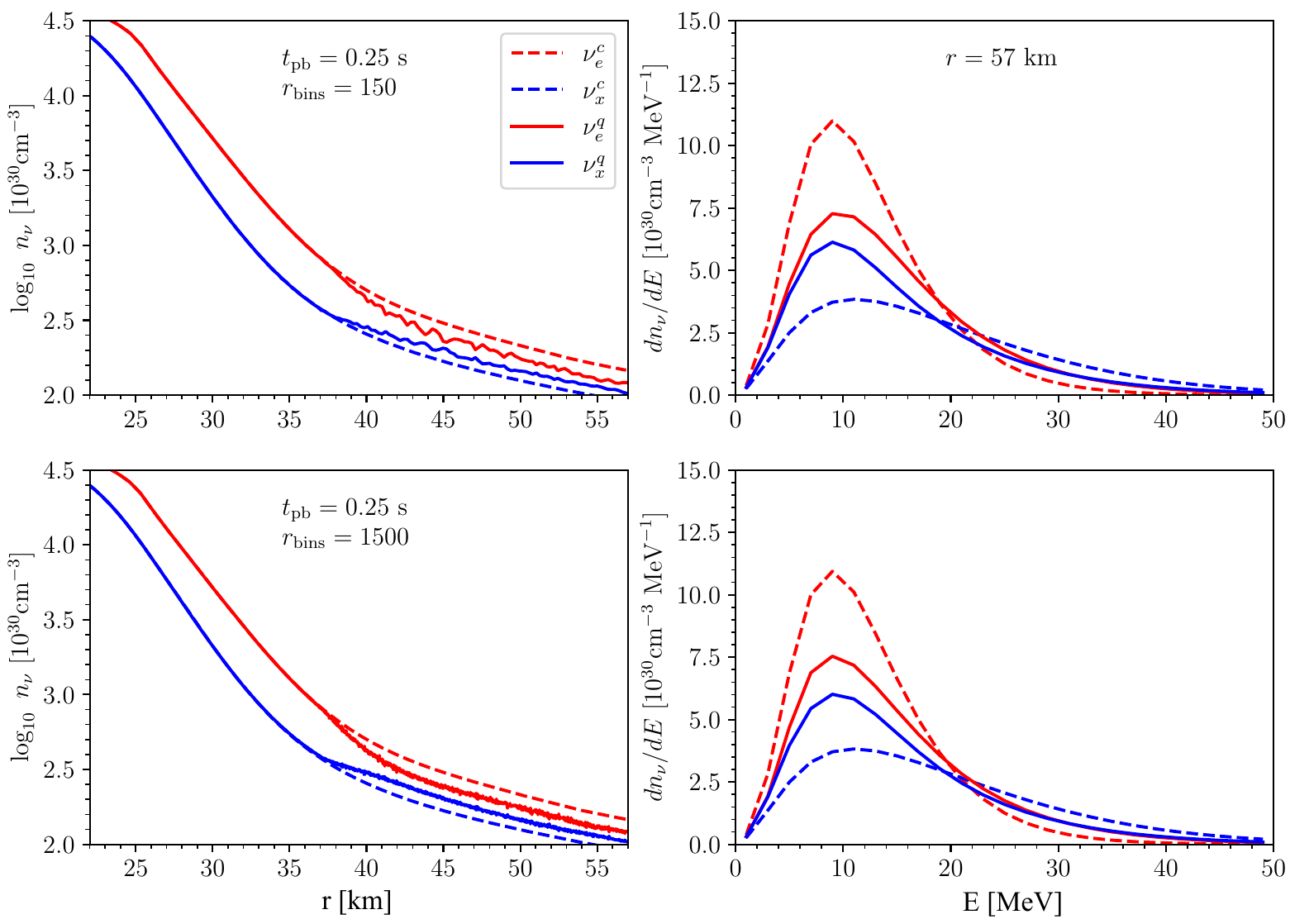}
		\includegraphics[width=0.90\textwidth]{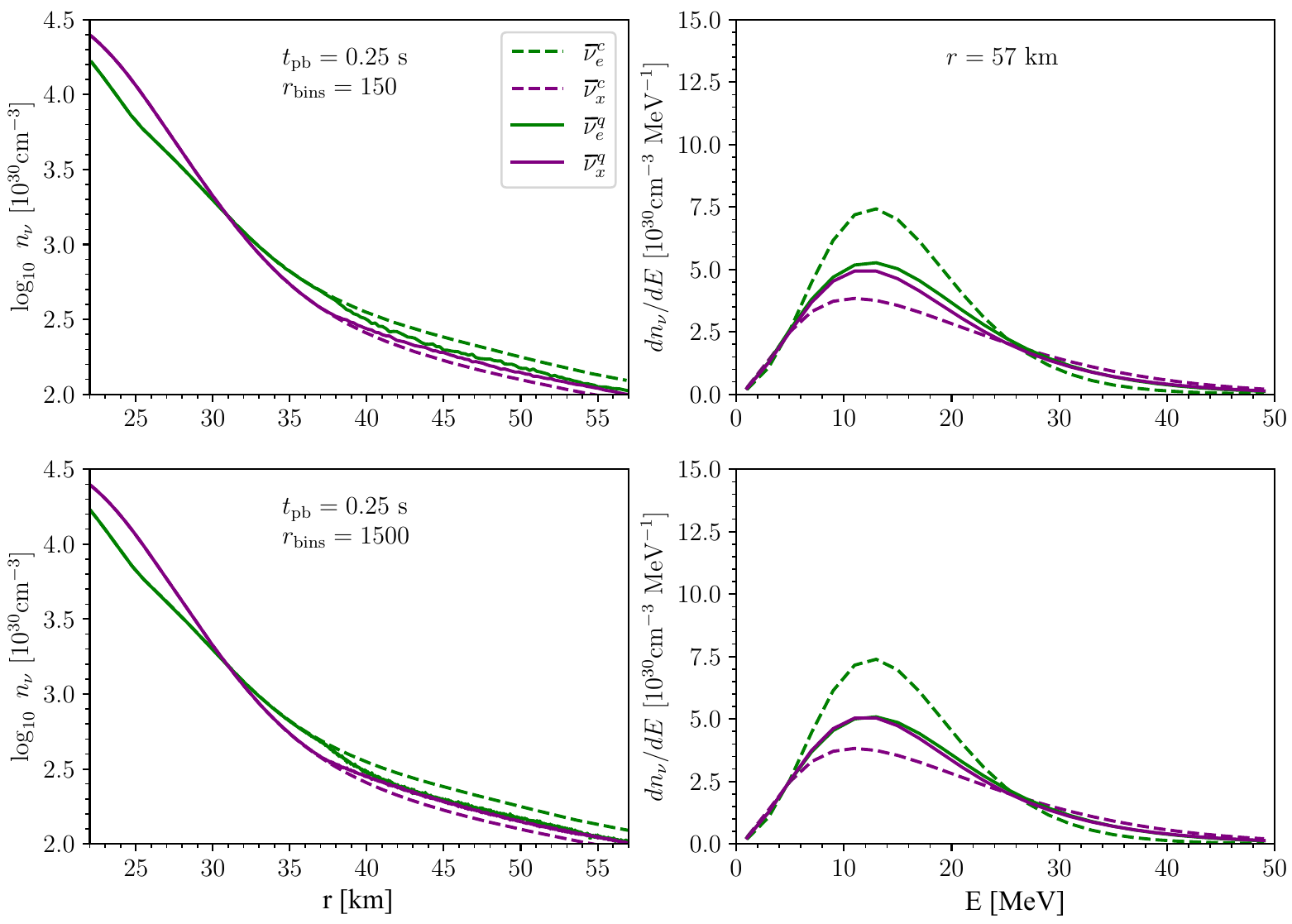}
		\caption{{\it Left panels:} Number densities of $\nu_e$'s and $\nu_x$'s as  functions of radius for $t_{\textrm{pb}} = 0.25$~s. The top (bottom) two panels display the number densities for neutrinos (antineutrinos)  obtained using $150$ and $1500$ radial bins.  The dashed lines show the number densities in the absence of neutrino flavor evolution for reference. {\it Right panels:} The top (bottom) two panels show the energy spectra of $\nu_{e}$ and $\nu_{x}$ ($\bar\nu_{e}$ and $\bar\nu_{x}$) extracted at  $r = 57$~km as  functions of energy obtained using $150$ and $1500$ radials bins. 
        }
		\label{tp25num}
\end{figure}

\begin{figure}[h!]
        \includegraphics[width=0.90\textwidth]{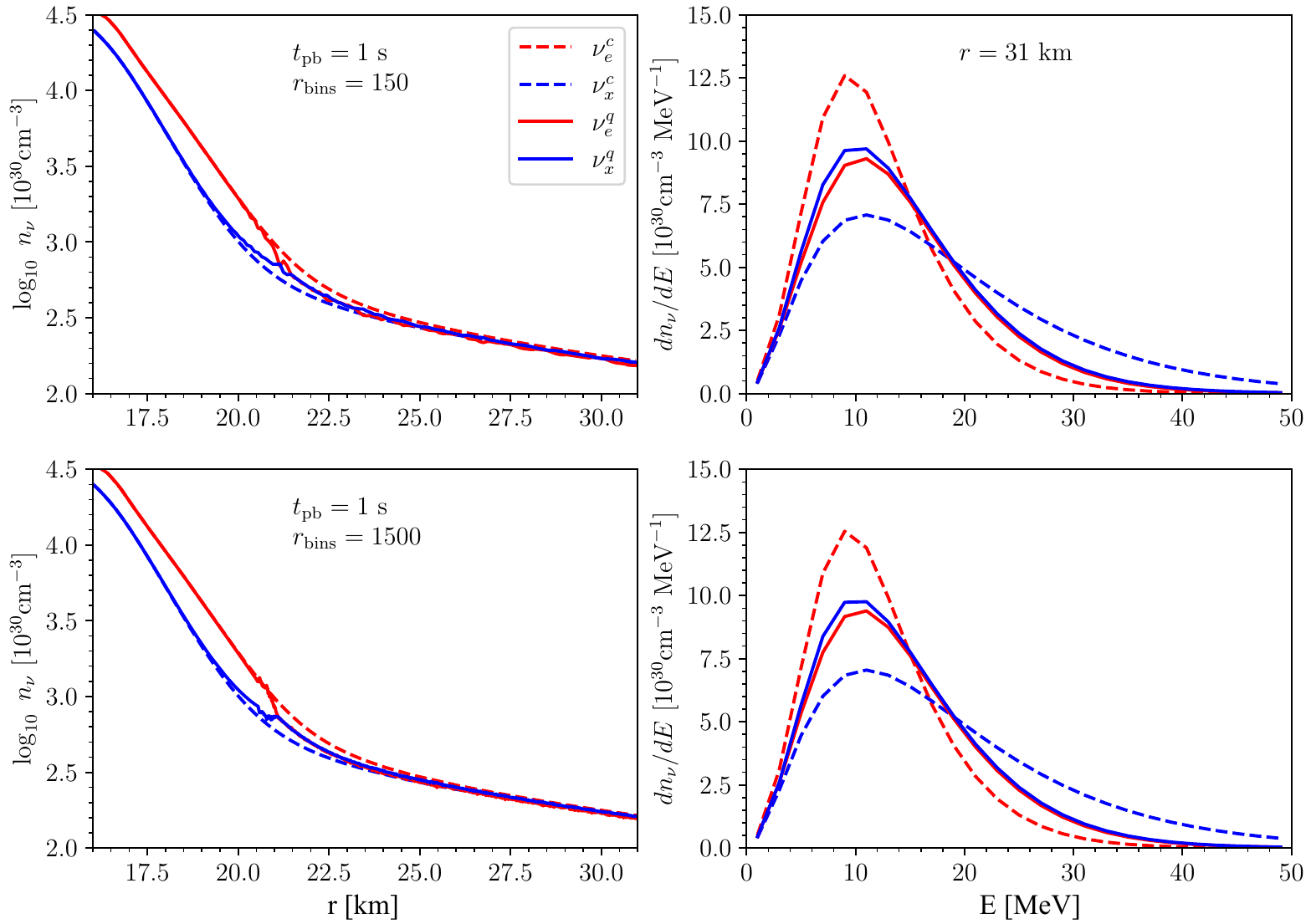}
        \includegraphics[width=0.90\textwidth]{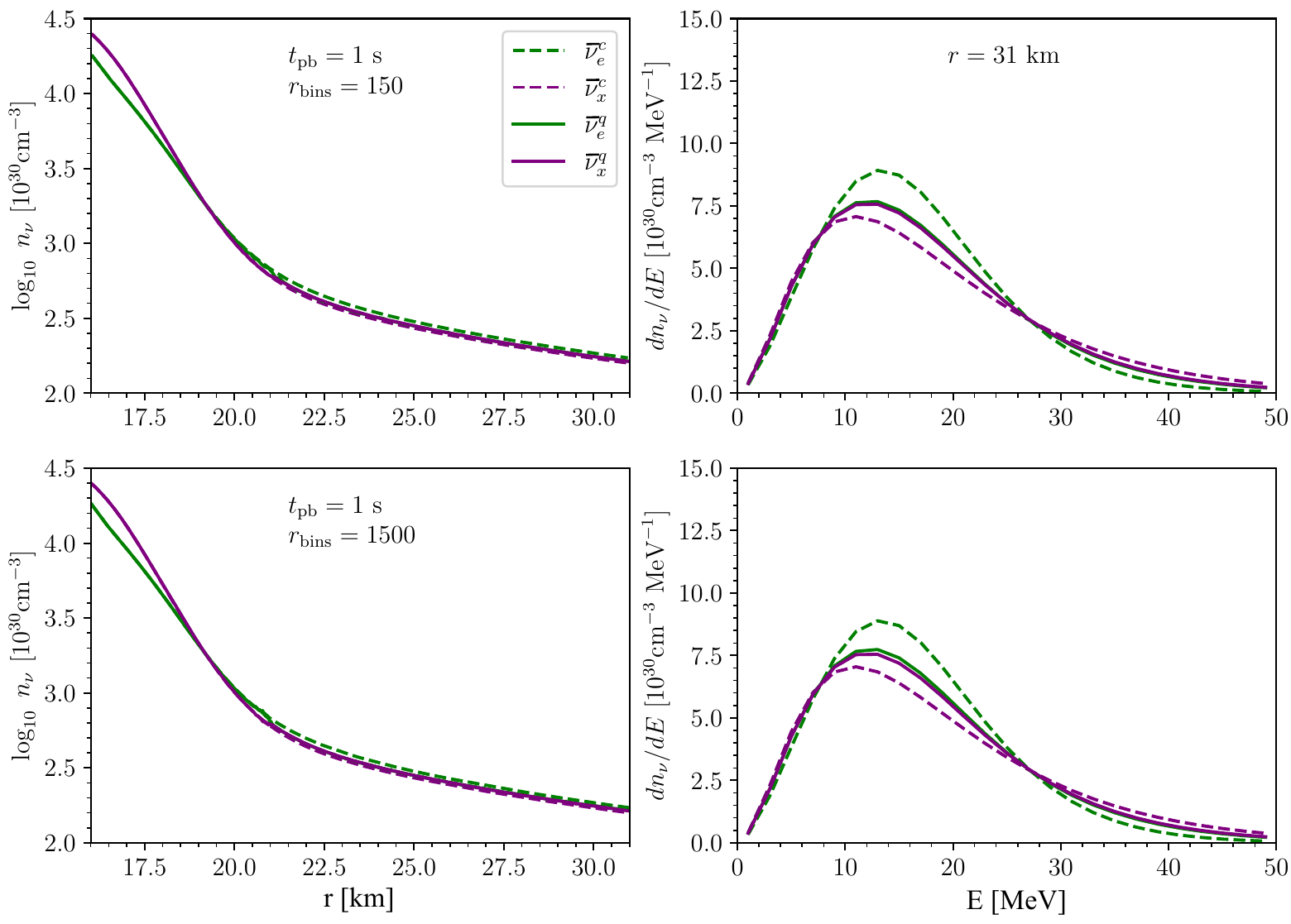}
		\caption{Same as Fig.~\ref{tp25num} but for $t_{\textrm{pb}} = 1$~s. The energy spectra are extracted at  $r=31$~km.}
        \label{t1num}
\end{figure}

Figures~\ref{tp25heat} and \ref{t1heat} show the quasi-steady state configurations obtained with $150$ and $1500$ radial bins (middle and bottom panels) for $0.25$~s and $1$~s, respectively. As it is clearly seen, the results obtained using $150$ radial bins are in excellent qualitative agreement with  the ones obtained using $1500$  bins, implying convergence. The  spots present in the quasi-steady-state configurations obtained  with $150$ radial bins, especially in the case of neutrinos,  can be attributed to fluctuations being more visible due to larger bin size. The convergence of the results with respect to the number of radial bins is more clearly observed in Figs.~\ref{tp25num} and \ref{t1num}, that show the radial evolution of the neutrino number densities (left panels) as  functions of radius  and  the energy spectra at $r_{\rm max}$ (right panels). These results demonstrate that the variation in number of radial bins  is only responsible for small quantitative variations, but it does not affect the overall physics outcome. 

Recently, Ref.~\cite{Nagakura:2025brr} explored flavor conversion in a simulation box with periodic boundary conditions, random perturbations, and neglecting collisions; they pointed out that  a coarser  spatial resolution may underestimate the growth of flavor instability, leading to wrong asymptotic states in the aftermath of flavor conversion.  However, we stress that the system explored in this paper is non-stationary and inhomogeneous, employs non-periodic boundaries, takes into account collisions with the background medium, and allows for neutrino decoupling on an extended spatial region. As pointed out in Refs.~\cite{Shalgar:2022rjj,Shalgar:2022lvv}, collisions and advection are crucial to smear small-scale structures and facilitate the employment of a coarser spatial grid, without loosing in accuracy in the numerical solution of the neutrino kinetic equations. In addition, the use of periodic boundary conditions may strongly affect the quasi-steady-state flavor configuration~\cite{Cornelius:2024zsb}. In the light of all these differences, the findings of Ref.~\cite{Nagakura:2025brr} cannot be applied to our system, but more work is needed  to verify the requirements of the numerical solution of the neutrino kinetic equations.


\bibliographystyle{JHEP}
\bibliography{bibliography.bib}
\end{document}